\documentclass[12pt]{article}
\usepackage{a4wide,amsmath,amsthm,epsfig,graphicx}
\usepackage{amsmath,amsthm,amssymb}
\usepackage{amsfonts}
\usepackage{graphicx}
\usepackage{latexsym}
\usepackage{bm}
\usepackage{algorithm}
\usepackage{algorithmic}
\usepackage{verbatim}

\usepackage[mathcal]{eucal}

\newcommand{\Qx}{ \mathbb{Q} }

\newcommand{\rec}{\mbox{R{\tiny EC}}}
\newcommand{\recinv}{\mbox{R{\tiny EC,0}}}
\newcommand{\reccou}{\mbox{R{\tiny EC,2}}}

\newcommand{\cash}{\mbox{C{\tiny ASHFLOWS}}}
\newcommand{\lgd}{\mbox{L{\tiny GD}}}

\newcommand{\npv}{\mbox{NPV}}

\newcommand{\cds}{\mbox{CDS}}
\newcommand{\brcva}{\mathrm{BR\text{-}CVA}}
\newcommand{\brcvacds}{\mathrm{BR\text{-}CVA\text{-}CDS}}

\newtheorem{theorem}{Theorem}[section]

\newtheorem{proposition}[theorem]{Proposition}
\newtheorem{remark}[theorem]{Remark}
\newtheorem{lemma}[theorem]{Lemma}

\newcommand{\LGD}[1]  {\mathrm{LGD}_{#1}}
\newcommand{\TLGD}[3] {\mathrm{LGD}_{#1,#2,#3}}

\newcommand{\CalculateAdjustment}{\mbox{\sc Calculate Adjustment }}
\newcommand{\CDSAdjust}{\mbox{\sc CDSAdjust}}
\newcommand{\ComputeProb}{\mbox{\sc ComputeProb}}

\numberwithin{equation}{section}

\title{%{\small This paper is available at www.damianobrigo.it  \\ Posted at ssrn.com in September 2006 }\\ \vspace{1cm}
\vspace{-1cm}
\bf{\Large Bilateral counterparty risk valuation\\ with stochastic
  dynamical models\\ and application to Credit Default Swaps}}
\author{ Damiano Brigo \\
Department of Mathematics, Imperial College, and \\
Fitch Solutions, 101 Finsbury Pavement, EC2A 1RS London, UK \\
E-mail: {\tt d.brigo@imperial.ac.uk}\\ \\
Agostino Capponi \\
California Institute of Technology \\
Division of Engineering and Applied Sciences \\
MC 256-80, 91125, Pasadena, CA, US \\
Email: {\tt acapponi@caltech.edu}\\\\
}

\date{\small First Version: Sept 23, 2008. This Version: \today}

\begin{document}

\maketitle
\thispagestyle{empty}

\begin{abstract}
We introduce the general arbitrage-free valuation framework for counterparty risk adjustments in presence of bilateral default risk, including default of the investor. We illustrate the symmetry in the valuation and show that the adjustment involves a long position in a put option plus a short position in a call option, both with zero strike and written on the residual net value of the contract at the relevant default times. We allow for correlation between the default times of the investor, counterparty and underlying portfolio risk factors. We use arbitrage-free stochastic dynamical models. We then specialize our analysis to Credit Default Swaps (CDS) as underlying portfolio, generalizing the work of Brigo and Chourdakis (2008)
\cite{Brigo08} who deal with unilateral and asymmetric counterparty risk.
We introduce stochastic intensity models and a trivariate copula function on the default times exponential variables to model default dependence.  Similarly to
\cite{Brigo08}, we find that both default correlation and credit spread
volatilities have a relevant and structured impact on the adjustment.
Differently from \cite{Brigo08}, the two parties will now agree on the credit valuation adjustment.
We study a case involving British Airways, Lehman Brothers
and Royal Dutch Shell, illustrating the bilateral adjustments in concrete crisis situations.

\end{abstract}

{\bf{AMS Classification Codes}}: 60H10, 60J60, 60J75, 62H20, 91B70 \\
{\bf{JEL Classification Codes}}: C15, C63, C65, G12, G13 \\

\noindent {\bf{keywords}}: Counterparty Risk, Arbitrage-Free Credit Valuation Adjustment, Credit
  Default Swaps, Contingent Credit Default Swaps, Credit Spread
  Volatility, Default Correlation, Stochastic Intensity, Copula
  Functions, Wrong Way Risk.

\pagestyle{myheadings}
\markboth{}{{\footnotesize  D. Brigo and A. Capponi, Arbitrage free bilateral counterparty risk valuation and application to CDS}}

\newpage

%\tableofcontents
%OB

\section{Introduction}

\subsection*{General motivation.}
In general, the reason to introduce counterparty risk when evaluating a contract is linked to the fact that many financial contracts are traded over the counter, so that the credit quality of the counterparty and of the investor may be important. This has become even more relevant in recent years as some protection sellers such as mono-line insurers or investment banks have witnessed increasing default probabilities or even default events, the case of Lehman Brothers being a clear example.

When investing in default risky assets, one requires a risk premium as a reward for assuming the default risk. If one thinks, for example, of a corporate bond, it is known that the yield is higher than the corresponding yield of an equivalent treasury bond, and this difference is usually called credit spread \cite{Brigo06}. The (positive) credit spread implies a lower price for the bond when compared to default free bonds. Many works have been proposed recently to explain the term structure of credit spreads, such as \cite{BrigoAlf05}, \cite{BrigoBachir08}, \cite{Duffie}, \cite{Coculescu} and \cite{BieleckiJeanblanc}, who also focuses on hedging.
This reduction in value is a typical feature: the value of a generic claim traded with a counterparty
subject to default risk  is always smaller than the value of the same
claim traded with a counterparty having a null default probability, as was shown formally for example in
\cite{BrigoMas}.

\subsection*{Bilateral risk and symmetry.}
This paper introduces a general arbitrage-free valuation framework for bilateral counterparty default risk.
By `bilateral' we intend to point out that the default of the investor is included into the framework, contrary to earlier works. This brings about symmetry, so that the price of the position including counterparty risk to the investor is exactly the opposite of the price of the position to the counterparty. This is clearly not the case if each of the two parties computes the present value assuming itself to be default-free and allowing for default of the other party only. This asymmetry would not matter in situations where financial investors had high credit quality and counterparties rather low one. Indeed, in such a case both parties would consider the investor as default-free and the counterparty as defaultable, so that inclusion of the investor default would be pointless, given that it happens in almost no scenario. However, recent events show that it is no longer realistic to take the credit quality of the financial institution for granted and to be highly superior to that of a general counterparty, no matter how prestigious or important the financial institution.

Bilateral risk is also mentioned in the credit risk measurement space by the Basel II documentation, Annex IV, 2/A:  ``Unlike a firm's exposure to credit risk through a loan, where the exposure to credit risk is unilateral and only the lending bank faces the risk of loss, the counterparty credit risk creates a bilateral risk of loss: the market value of the transaction can be positive or negative to either counterparty to the transaction."

Basel II is more concerned with Risk Measurement than pricing. For an analysis of Counterparty risk in the risk measurement space we refer for example to De Prisco and Rosen (2005) \cite{deprisco}, who consider modeling of stochastic credit exposures for derivatives portfolios.  However, also in the valuation space, bilateral features are quite relevant and often can be responsible for seemingly paradoxical statements.\footnote{We are grateful to Dan Rosen for first signaling this issue to us during a conference in June 2009} For example, Citigroup in its press release on the first quarter revenues of 2009 reported a {\em positive} mark to market due to its {\em worsened} credit quality: ``Revenues also included [...] a net 2.5USD billion positive CVA on derivative positions, excluding monolines, mainly due to the widening of Citi's CDS spreads". In this paper we explain precisely how such a situation may origin.

This paper first generalizes to the bilateral symmetric case the asymmetric unilateral setting proposed in \cite{Brigo08},
\cite{BrigoChourBakkar}, \cite{BrigoMas} and \cite{BrigoPall}, where only the counterparty is subject to default risk, while the investor is assumed to be default free.
We provide a general formula that gives the bilateral risk credit valuation adjustment (BR-CVA) for portfolios exchanged between a default risky investor and a default risky counterparty. Such formula shows that the adjustment to the investor is the difference between two discounted
options terms, a discounted call option in scenarios of early default time of the counterparty minus a discounted put option in scenarios of early default times of the investor, both options being on the residual net present value of the portfolio at the relevant default times and having zero strike. The BR-CVA seen from the point of view of the counterparty is exactly the opposite.
We allow for correlation between default of the investor, default of the counterparty and underlying portfolio risk factors, and for volatilities and dynamics in the credit spreads and in the underlying portfolio, all arbitrage free.

\subsection*{Bilateral Counterparty adjustment applied to CDS.}
We then specialize our analysis to Credit Default Swaps (CDS) as underlying portfolio, generalizing the work of Brigo and Chourdakis
\cite{Brigo08} who deal with unilateral and asymmetric counterparty risk for these contracts. Featuring a CDS as underlying, a third default time enters the picture, namely the default time for the reference credit of the CDS. We therefore assume that all three entities are subject to default risk and that the default events of investor, counterparty and reference credit are correlated. We then propose a numerical methodology to evaluate the resulting BR-CVA formula. We investigate the impact of both credit spread volatility and default
correlation on the credit valuation adjustment. Most of previous approaches on CDS counterparty risk only focus on unilateral
counterparty risk, moreover ignoring the effect of volatility on the adjustment
and mainly focusing on default correlation.

The few earlier works on CDS with counterparty risk include Leung and Kwok (2005)\cite{Leung} who, building on Collin-Dufresne et al. (2002) \cite{Collin2002}, model default intensities as deterministic constants with default indicators of other names as feeds. The exponential triggers of the default times are taken to be independent and default correlation results from the cross feeds, although again there is no explicit modeling of credit spread volatility. Hull and White \cite{Hull2000} study the counterparty risk problem by resorting to barrier correlated models. Walker (2005) \cite{Walker} models CDS counterparty risk using transitions rates as natural means to represent contagion, but again ignores credit spread volatility.  Hille et al. (2006)\cite{hille} concentrate on credit risk measurement for CDS rather than precise valuation under counterparty risk.

The need for explicitly modeling credit spread volatility is even more pronounced if the underlying reference contract is itself a CDS, as the credit valuation adjustment would involve CDS options and it is very undesirable to model options without volatility in the underlying asset. This has been clearly shown in \cite{Brigo08} for CDS, and in \cite{BrigoPall} for interest rates payoffs, and again in \cite{BrigoChourBakkar} for commodities payoffs.
Indeed, most credit models in the industry, especially when applied to Collateralized Debt Obligations or $k$-th to default baskets, model default correlation but ignore credit spread volatility. Credit spreads are typically assumed to be deterministic and a copula is postulated on the exponential triggers of the default times to model default correlation. This is the opposite of what used to happen with counterparty risk for interest rate underlyings, for example in Sorensen and Bollier (1994) \cite{Sorensen} or Brigo and Masetti \cite{BrigoMas} (in Pykhtin (2006)\cite{Pyk}), where correlation was ignored and volatility was modeled instead. Brigo and Chourdakis (2008) \cite{Brigo08} rectify this when addressing CDS's, but only deal with unilateral and asymmetric counterparty risk.

Summarizing, we can put this paper's contribution into context with respect to its analogous earlier versions for unilateral counterparty risk or for other asset classes through Table~\ref{table:contextCRmodels}.

Other recent works include Cr\'epey et al (2009)\cite{crepey} who model wrong way risk for CDS with counterparty risk using a Markov chain copula model, and  Blanchet-Scalliet and Patras (2008) \cite{patras}, who resort to older Merton type models. A structural model with jumps has been introduced by Lipton and Sepp (2009) \cite{lipton}. In these models assumptions on credit spread volatility are most time implicit.

Here instead we introduce stochastic intensity models for all three names, including the investor. We do not introduce correlation between the three intensity processes, but model default correlation through a
trivariate copula function on the exponential triggers of the default times.
This is because typically spread correlation has a much lower impact on dependence of default times than default correlation. The correlation structure underlying the trivariate copula function may be estimated by the implied correlation in the quoted indices tranches markets such as i-Traxx and CDX or calculated from the asset correlation of the names. We leave this problem for further research. We present a preliminary numerical investigation to highlight the impact of dynamics parameters on bilateral CVA, followed by a case study of bilateral risk based on British Airways, Lehman Brothers and Royal Dutch Shell. We confirm findings in \cite{Brigo08} showing that both default correlation and credit spread volatilities have a relevant and structured impact on the adjustment.

We specify that we do not consider specific collateral clauses or guarantees in the present work. We assume we are dealing with counterparty risk for an over the counter CDS transaction where there is no periodic margining or collateral posting. The aim of this paper is analyzing the fine structure of counterparty risk adjustments with respect to finely tuned market dynamics, including the fundamental credit spread volatility ignored in other approaches, and wrong way risk. We plan to address collateral provisions inclusion in future work. We already addressed netting in the interest rate context in \cite{BrigoPall} and \cite{BrigoMas}. The impact of credit triggers for the counterparty on CVA are analyzed in Yi (2009) \cite{Yi}. Assefa et al (2009) \cite{assefa}  analyze the modeling of collateralization and margining in CVA calculations.
\begin{table}[h!]
\begin{tabular}{|c|c|c|c|c|}
\hline
      Modeling $\rightarrow$   & Underlying Volatility  & Underlying Vol + &  Bilateral \\
 Asset Class  $\downarrow$         & No underl/counterp Corr &  underl/counterp Corr & + Vol and Corr \\
         \hline
IR swaps& (Net) B.  Masetti \cite{BrigoMas}  &             (Net)    B. Pallavicini \cite{BrigoPall}  & (Net) Brigo Pallavicini \\
        &   &                                         &   \& Papatheodorou \cite{BrigoPallPap}         \\
                &   &                                         &   (forthcoming)         \\
IR exotics&                         &               B. Pallavicini \cite{BrigoPall} & B. P. P. \cite{BrigoPallPap}   \\ & & & (forthcoming) \\
Oil swaps &                         &              B.  Bakkar \cite{BrigoChourBakkar}   &          \\ & & & \\
CDS       &                         &  B.  Chourdakis \cite{Brigo08} & {\bf This paper} \\ & & & \\
Equity    &                         &  B.  Masetti \cite{BrigoMas}  &                        \\ & & & \\ \hline
\end{tabular}
\caption{Part of earlier analogous literature on CVA valuation with respect to inclusion of volatilities and correlation and of bilateral features. ``Net" denotes papers that consider netting.}\label{table:contextCRmodels}
\end{table}

\subsection*{Paper structure.}

The rest of the paper is organized as follows. Section \ref{sec:generalcounterformula} gives the general BR-CVA
formula and refers to the appendix for the mathematical proof.
Section \ref{sec:CDS_application} defines the underlying framework
needed for applying the above methodology to credit default swaps,
including the stochastic intensity model and the trivariate copula
function for correlating defaults. Section
\ref{sec:numerical_algorithm} develops a numerical method to
implement the formula in the case when the underlying contract is a
CDS and gives the pseudo code of the algorithm used to run
numerical simulations.
Section \ref{sec:case_study} present numerical experiments on
different default correlation scenarios and changes in credit spread
volatility. Section \ref{sec:LehmanExperiment} presents a specific application
for computing the mark-to-market value of a CDS contract agreed
between British Airways, Lehman Brothers and Royal Shell.
 Section \ref{sec:conclusion} concludes the paper.

\section{Arbitrage-free valuation of bilateral counterparty risk}\label{sec:generalcounterformula}
We refer to the two names involved in the financial contract and subject to default risk as
\begin{eqnarray}
\nonumber \text{investor } & \rightarrow & \text{name ``0''} \\
\nonumber \text{counterparty } & \rightarrow & \text{name ``2''}
\label{eq:three_names}
\end{eqnarray}

In cases where the portfolio exchanged by the two parties is also a default sensitive instrument, we introduce a third name referring to the reference credit of that portfolio

\[ \text{reference credit }  \rightarrow  \text{name ``1''} \]

If the portfolio is not default sensitive then name ``1'' can be removed.

We follow \cite{BrigoBudapest} and denote by $\tau_0$, ($\tau_1$) and
$\tau_2$ respectively the default times of the investor, (reference credit) and counterparty.
We place ourselves in a probability space $(\Omega,\mathcal{G},\mathcal{G}_t,\mathbb{Q})$.
The filtration $\mathcal{G}_t$ models the flow of information
of the whole market, including credit and $\Qx$ is the risk neutral
measure. This space is endowed also with a right-continuous and
complete sub-filtration $\mathcal{F}_t$ representing all the
observable market quantities but the default event, thus
$\mathcal{F}_t\subseteq\mathcal{G}_t:=\mathcal{F}_t\vee\mathcal{H}_t$.
 Here, $\mathcal{H}_t=\sigma(\{\tau_0\leq u \} (\vee \{\tau_1\leq u \}) \vee \{\tau_2\leq u\} :u\leq t)$ is
the right-continuous filtration generated by the default events,
either of the investor or of his counterparty (and of the reference credit if the underlying portfolio is credit sensitive).
We also introduce the notion of stopped filtration. If $\tau$ is an $\mathcal{F}_t$ stopping time, then the stopped filtration
$\mathcal{F}_{\tau}$ is defined as
\begin{eqnarray}
\mathcal{F}_{\tau} = \sigma(\mathcal{F}_t \cup \{t \leq \tau \}, t \geq 0)
\end{eqnarray}
If $\tau$ is an $\mathcal{G}_t$ stopping time, then the stopped filtration
$\mathcal{G}_{\tau}$ is defined as
\begin{eqnarray}
\mathcal{G}_{\tau} = \sigma(\mathcal{G}_t \cup \{t \leq \tau \}, t \geq 0)
\end{eqnarray}

Let us call $T$ the final maturity of the payoff which we need to
evaluate and let us define the stopping time

\begin{equation}
\tau = \min\{\tau_0, \tau_2\}
\label{eq:stopping_time}
\end{equation}

If $\tau>T$, there is neither default of the investor, nor of his counterparty
during the life of the contract and they both can fulfill the
agreements of the contract. On the contrary, if $\tau\leq
T$ then either the investor or his counterparty (or both) defaults.
 At $\tau$, the Net Present Value (NPV) of the
residual payoff until maturity is computed. We then distinguish two
cases:
\begin{itemize}
\item $\tau = \tau_2$. If the NPV is negative (respectively positive) for the investor (defaulted
counterparty), it is completely paid (received) by the investor
(defaulted counterparty) itself. If the NPV is positive (negative) for the
investor (counterparty), only a recovery fraction $\reccou$ of the
NPV is exchanged.
\item $\tau = \tau_0$. If the NPV is positive (respectively negative)
  for the defaulted investor (counterparty), it is completely received
  (paid) by the defaulted investor
(counterparty) itself. If the NPV is negative (positive) for the
defaulted investor (counterparty), only a recovery fraction $\recinv$ of the
NPV is exchanged.
\end{itemize}
Let us define the following (mutually exclusive and exhaustive) events ordering the default times

\begin{eqnarray}
\nonumber A &=& \{\tau_0 \le \tau_2 \le T\} \ \ \ \ \  E = \{T \le \tau_0 \le \tau_2 \} \\
\nonumber B &=& \{\tau_0 \le T \le \tau_2 \} \ \ \ \ \    F = \{T \le \tau_2 \le \tau_0 \} \\
\nonumber C &=& \{\tau_2 \le \tau_0 \le T \}\\
D &=& \{\tau_2 \le T \le \tau_0 \}
\label{eq:event_set}
\end{eqnarray}

Let us call $\Pi^D(t,T)$ the discounted payoff of a generic defaultable claim at $t$ and
$\cash(u,s)$ the net cash flows of the claim without default
between time $u$ and time $s$, discounted back at $u$, all payoffs
seen from the point of view of the investor.
Then, we have
$\npv(\tau_i)=\mathbb{E}_{\tau_i}\{\cash(\tau_i,T)\}$, $i=0,2$. Let us denote by $D(t,T)$ the price of a
zero coupon bond with maturity $T$. We have
\begin{eqnarray}\label{generalpayoff}
\nonumber \Pi^D(t,T) & = & \mathbf{1}_{E \cup F} \cash(t,T) \\
\nonumber & &+ \mathbf{1}_{C \cup D}
\left[\cash(t,\tau_2)+D(t,\tau_2)\left(\reccou\left(\npv(\tau_2)\right)^+-\left(-\npv(\tau_2)\right)^+\right)\right] \\
& & + \mathbf{1}_{A \cup B}
\nonumber \left[\cash(t,\tau_0)+D(t,\tau_0)\left(\left(\npv(\tau_0)\right)^+
  -\recinv\left(-\npv(\tau_0)\right)^+\right)\right] \\
\end{eqnarray}
This last expression is the general payoff under
bilateral counterparty default risk. Indeed, if there is no early
default, this expression reduces to risk neutral valuation of the
payoff (first term on the right hand side). In case of early
default of the counterparty, the payments due before default occurs are received
(second term), and then if the residual net present value is
positive only the recovery value of the counterparty $\reccou$ is received (third term), whereas
if it is negative it is paid in full by the investor (fourth term). In case of early
default of the investor, the payments due before default occurs are received
(fifth term), and then if the residual net present value is
positive it is paid in full by the counterparty to the investor (sixth
term), whereas if it is negative only the recovery value of the
investor $\recinv$ is paid to the counterparty (seventh term).

Let us denote by $\Pi(t,T)$ the discounted payoff for an equivalent claim
with a default-free counterparty, i.e. $\Pi(t,T) = \cash(t,T)$. We then have
 the following
\begin{proposition} [\cite{BrigoBudapest}] {\bf (General bilateral counterparty risk pricing
formula)}\label{prop_main}. At valuation time $t$, and conditional on the event $\{
  \tau > t\}$, the price of the payoff under bilateral counterparty risk is
\begin{eqnarray}\label{generalprice}
\nonumber \mathbb{E}_t \left\{ \Pi^D(t,T)\right\}& = &
\mathbb{E}_t\left\{ \Pi(t,T)\right\}\\
\nonumber & & +  \mathbb{E}_t\left\{\LGD{0} \cdot \mathbf{1}_{A \cup B}  \cdot D(t,\tau_0) \cdot \left[ - \mathrm{NPV}(\tau_0)\right]^+\right\}\\
 & & -   \mathbb{E}_t\left\{\LGD{2} \cdot \mathbf{1}_{C \cup D} \cdot D(t,\tau_2) \cdot \left[ \mathrm{NPV}(\tau_2)\right]^+\right\}
\end{eqnarray}
where $LGD =1-\rec$ is the \emph{Loss Given Default} and the
recovery fraction $\rec$ can be stochastic and possibly correlated with the default indicator process.
It is clear that the value of a defaultable claim is the value of the
corresponding default-free claim plus a long position in a put option
(with zero strike) on the residual NPV giving nonzero contribution
only in scenarios where the investor is the earliest to default (and does so before final maturity) plus
a short position in a call option (with zero strike) on the residual
NPV giving nonzero contribution in scenarios where the counterparty is
the earliest to default (and does so before final maturity),
\end{proposition}
Proposition \ref{prop_main} is stated in \cite{BrigoBudapest} without
a proof. Here, we provide a mathematical proof in Appendix \ref{sec:proof_prop}.
The adjustment is called bilateral counterparty risk credit valuation
adjustment (BR-CVA) and it may be either positive or negative
depending on whether the counterparty is more or less likely to
default than the investor and on the volatilities and correlation.
The mathematical expression is given by
\begin{eqnarray}
\nonumber \brcva(t,T,\TLGD{0}{1}{2}) &=&  \mathbb{E}_t\left\{
 \LGD{2} \cdot \mathbf{1}_{C \cup D} \cdot D(t,\tau_2) \cdot \left[
 \mathrm{NPV}(\tau_2)\right]^+\right\} \\
 & & - \mathbb{E}_t\left\{\LGD{0} \cdot \mathbf{1}_{A \cup B}  \cdot D(t,\tau_0) \cdot \left[ - \mathrm{NPV}(\tau_0)\right]^+\right\}
\label{eq:br-cva}
\end{eqnarray}
where the right hand side in Eq. (\ref{eq:br-cva}) depends on $T$
through the events $A,B,C,D$ and $\LGD{012} = (\LGD{0},\LGD{1},\LGD{2})$
is shorthand notation to denote the dependence on
the loss given defaults of the three names.

\begin{remark} {\bf (Symmetry vs Asymmetry).} With respect to earlier results on counterparty risk valuation,
Equation (\ref{eq:br-cva}) has the great advantage of being symmetric. This is to say that if ``2" were to compute counterparty risk of her position towards ``1", she would find exactly $-\brcva(t,T,\TLGD{0}{1}{2})$. However, if each party computed the adjustment by assuming itself to be default-free and considering only the default of the other party, then the adjustment calculated by ``0'' would be
\[ \mathbb{E}_t\left\{\LGD{2} \cdot
 \mathbf{1}_{\tau_2 < T} \cdot D(t,\tau_2) \cdot \left[
 \mathrm{NPV}(\tau_2)\right]^+\right\}
\]
whereas the adjustment calculated by ``2'' would be
\[  \mathbb{E}_t\left\{\LGD{0} \cdot \mathbf{1}_{\tau_0 < T}  \cdot D(t,\tau_0) \cdot \left[ - \mathrm{NPV}(\tau_0)\right]^+\right\} \]
and they would not be one the opposite of the other. This means that only in the first case the two parties agree on the value of the counterparty risk adjustment.
\end{remark}

\begin{remark} {\bf (Change in sign).}
Earlier results on counterparty risk valuation, concerned with a default-free investor and asymmetric, would find an adjustment to be subtracted that is always positive. However, in our symmetric case even if the initial adjustment is positive due to
 \[\mathbb{E}_t \left\{\LGD{2} \cdot
 \mathbf{1}_{C \cup D} \cdot D(t,\tau_2) \cdot \left[
 \mathrm{NPV}(\tau_2)\right]^+\right\} >
  \mathbb{E}_t\left\{\LGD{0} \cdot \mathbf{1}_{A \cup B}  \cdot D(t,\tau_0) \cdot \left[ - \mathrm{NPV}(\tau_0)\right]^+\right\}\]
the situation may change in time, to the point that the two terms may cancel or that the adjustment may change sign as the credit quality of ``0" deteriorates and that of ``2" improves, so that the inequality changes direction.
\end{remark}

\begin{remark} {\bf (Worsening of credit quality and positive mark to market).}
If the Investor marks to market her position at a later time using Formula (\ref{generalprice}), we can see that the term in $\lgd_0$ increases, ceteris paribus, if the credit quality of ``0" worsens. Indeed, if we increase the credit spreads of the investor, now $\tau_0 < \tau_2$ will happen more often, giving more weight to the term in $\lgd_0$. This is at the basis of statements like the above one of Citigroup.
\end{remark}

\section{Application to Credit Default Swaps} \label{sec:CDS_application}
In this section we use the formula developed in Section
\ref{sec:generalcounterformula} to evaluate the BR-CVA in credit
default swap contracts (CDS). Subsection \ref{subs:CDSpayoff} recalls
the general formula for CDS evaluation. Subsection \ref{subs:copula}
introduces the copula models used to correlate the default
events. Subsection \ref{subs:intensity} recalls the CIR model used for
the stochastic intensity of the three names. Subsection
\ref{subs:CDSadj} applies the general BR-CVA formula to
calculate the adjustment for CDS contracts. We restrict our attention to CDS
contract without an upfront trading. However, the proposed methodology is perfectly applicable to the case when the CDS trades with an upfront premium, which has become the case after the big bang protocol.

\subsection{CDS Payoff}\label{subs:CDSpayoff}
 We assume deterministic interest rates, which
leads to independence between $\tau_1$ and $D(0,t)$,
and deterministic recovery rates.
Our results hold also true for the case of stochastic rates
independent of default times.
The receiver CDS valuation, for a CDS selling protection LGD$_1$ at time 0
for default of the reference entity between times $T_a$ and $T_b$ in exchange of a periodic premium rate $S_1$ is given by
\begin{eqnarray}
\nonumber \cds_{a,b}(0, S_1, \LGD{1}) &=& S_1
\bigg[-\int_{T_a}^{T_b}D(0,t) (t-T_{\gamma(t)-1}) d\mathbb{Q}(\tau_1 > t) \\
\nonumber & & + \sum_{i=a+1}^b \alpha_i D(0,T_i) \mathbb{Q}(\tau_1 > T_i)
\bigg]  \\
 & & +\LGD{1} \left[\int_{T_a}^{T_b} D(0,t) d\mathbb{Q}(\tau_1 > t) \right]
\label{eq:cds_payoff}
\end{eqnarray}
where $\gamma(t)$ is the first payment period $T_j$ following time
$t$. Let us denote by
\begin{equation}
\npv(T_j,T_b):= \cds_{a,b}(T_j, S, \LGD{1})
\label{eq:cds_payoff_interm}
\end{equation}
the residual NPV of a receiver CDS between $T_a$ and $T_b$ evaluated
at time $T_j$, with $T_a < T_j < T_b$. Eq. (\ref{eq:cds_payoff_interm}) can
be written similarly to Eq. (\ref{eq:cds_payoff}), except that now
evaluation occurs at time $T_j$ and has to be conditioned on the
information set available to the market at $T_j$. This leads to

\begin{eqnarray}
\nonumber \cds_{a,b}(T_j, S_1, \LGD{1}) &=&
\mathbf{1}_{\tau_1 > T_j} \overline{\cds}_{a,b}(T_j,S_1,LGD_1) \\
\nonumber &:=& \mathbf{1}_{\tau_1 > T_j} \bigg\{ S_1
\bigg[-\int^{T_b}_{\max\{T_a,T_j\}} D(T_j,t) (t-T_{\gamma(t)-1})
d \mathbb{Q}(\tau_1 > t | \mathcal{G}_{T_j}) \\
\nonumber & & + \sum_{i = \max\{a,j\}+1}^b \alpha_i D(T_j,T_i)
\mathbb{Q}(\tau_1 > T_i | \mathcal{G}_{T_j}) \bigg] \\
 & & + \LGD{1} \bigg[\int^{T_b}_{\max\{T_a,T_j\}} D(T_j,t)
  d\mathbb{Q}(\tau_1 > t | \mathcal{G}_{T_j}) \bigg]  \bigg\}
\label{eq:cds_payoff_interm_develop}
\end{eqnarray}

For conversion of these running CDS into upfront ones, following the so called Big Bang protocol by ISDA, see for example Beumee, Brigo, Schiemert and Stoyle (2009)\cite{beumee}. Our reasoning still applies with obvious modifications to the upfront CDS contract.

\subsection{Default Correlation}\label{subs:copula}
We consider a reduced form model that is stochastic in the default
intensity for the investor, counterparty and CDS reference credit.
The default correlation between the three names is defined through a
dependence structure on the exponential random variables characterizing the
default times of the three names. Such dependence structure is modeled using
 a trivariate copula function. Let us denote by $\lambda_i(t)$ and
$\Lambda_i(t) = \int_0^t \lambda_i(s) ds$ respectively the
default intensity and cumulated intensity of name $i$ evaluated at
time $t$. We recall that $i = 0$ refers to the investor, $i=1$ refers
to the reference credit and $i=2$ to the counterparty. We assume $\lambda_i$ to be independent of $\lambda_j$
for $i \neq j$, and assume each of them to be strictly positive almost
everywhere, thus implying that $\Lambda_i$ is invertible. We stress the fact that independence
of $\lambda$'s across names does not mean that the default event of one
name does not change the default probability or intensity of other
names, as discussed in \cite{Brigo08}, Section 4.
We place ourselves in a Cox process setting, where
\begin{equation}
\tau_i = \Lambda_i^{-1}(\xi_i), \ i = 0,1,2
\label{eq:default_times}
\end{equation}
with $\xi_0$, $\xi_1$ and $\xi_2$ being standard (unit-mean)
exponential random variables whose associated uniforms
\begin{equation}
U_i = 1 -\exp\{-\xi_i\}
\label{eq:unif_implied}
\end{equation}
 are correlated through a Gaussian trivariate copula
function
\begin{equation}
C_{\bm{R}}(u_0,u_1,u_2) = \mathbb{Q}(U_0 < u_0, U_1 < u_1, U_2 < u_2)
\label{eq:trivar_copula}
\end{equation}
with $\bm{R} = [r_{i,j}]_{i,j=0,1,2}$ being the correlation matrix
parameterizing the Gaussian copula.
Notice that a trivariate Gaussian copula implies bivariate Gaussian marginal
copulas. We show it for the case of the bivariate copula connecting the reference
credit and the counterparty, but the same argument applies to the
other bivariate copulas. Let $(X_1,X_2,X_3)$ be a standard Gaussian vector with correlation
matrix $\bm{R}$ and let $\Phi_{\bm{R}}$ be the distribution function of a
multivariate Gaussian random variable with correlation matrix $\bm{R}$. For
any pair of indices $i \neq j$, $0 \leq i,j \leq 2$, we denote by
$\bm{R}_{i,j}$ the $2 \cdot 2$ submatrix formed by the
intersection of row $i$ and row $j$ with column $i$ and column $j$.
We next state, without proof, an obvious result as a Lemma, which will be later used.
\begin{lemma}
A trivariate Gaussian copula with correlation matrix $\bm{R}$ induces
 marginal bivariate Gaussian copulas.
\end{lemma}
\noindent We denote by $C_{i,j}(u_i,u_j)$ the bivariate copula associated to $\bm{R}_{i,j}$.

\begin{comment}
\proof
We prove that the bivariate Gaussian copula $C_{1,2}(u_1,u_2)$
connecting the reference credit and the counterparty is obtained from
the trivariate Gaussian copula by taking the submatrix $\bm{R}$
formed by the intersection of rows 1 and 2 with columns 1 and 2.
We have
\begin{eqnarray}
\nonumber \mathbb{Q}(U_1 < u_1, U_2 < u_2) &=& \mathbb{Q}(U_1 < u_1, U_2 < u_2,
U_3 < 1) \\
\nonumber &=& C_{\bm{R}}(u_1,u_2,1) \\
\nonumber &=& \lim_{b \rightarrow 1^{-}} \Phi_{\bm{R}}(\Phi^{-1}(u_1), \Phi^{-1}(u_2), \Phi^{-1}(b)) \\
\nonumber &=& \lim_{b \rightarrow 1^{-}} \mathbb{Q}(X_1 <
\Phi^{-1}(u_1), X_2 < \Phi^{-1}(u_2),X_3 < \Phi^{-1}(b)) \\
\nonumber &=& \mathbb{Q}(X_1 < \Phi^{-1}(u_1), X_2 < \Phi^{-1}(u_2)) \cdot \\
\nonumber & & \lim_{b \rightarrow 1^{-}} \mathbb{Q}(X_3 < \Phi^{-1}(b) | X_1 < \Phi^{-1}(u_1), X_2 < \Phi^{-1}(u_2))\\
\nonumber &=& \Phi_{\bm{R}_{1,2}}(\Phi^{-1}(u_1),\Phi^{-1}(u_2))\\
          &:=& C_{1,2}(u_1,u_2)
\label{eq:biv_gauss_copula}
\end{eqnarray}
By a similar calculation, the bivariate copula connecting the investor and the reference credit is
given by $C_{0,1}$ with associated correlation matrix $\bm{R}_{0,1}$ and
the bivariate copula connecting the investor and the counterparty is
given by $C_{0,2}$ with associated correlation matrix $\bm{R}_{0,2}$.
\endproof
\end{comment}
\subsection{CIR stochastic intensity model}\label{subs:intensity}
We assume the following stochastic intensity model \cite{BrigoAlf05},
\cite{BrigoBachir08} for the three names
\begin{equation}
\lambda_j(t) = y_j(t) + \psi_j(t;\bm{\beta}_j), \ t \geq 0, j = 0,1,2
\end{equation}
where $\psi$ is a deterministic function, depending on the parameter
vector $\bm{\beta}$ (which includes $y_0$), that is integrable on closed
intervals. We assume each $y_j$ to be a Cox Ingersoll Ross (CIR) process
\cite{Brigo06} given by

\begin{equation}
dy_j(t) = \kappa_j(\mu_j-y_j(t))dt + \nu_j \sqrt{y_j(t)} dZ_j(t) +
J_{M_t,j} dM_j(t), j = 0,1,2
\end{equation}
where $J$'s are i.i.d. positive jump sizes that are exponentially
distributed with mean $\chi_j$ and $M_j$ are Poisson processes with
intensity $m_j$ measuring the arrival of jumps in the intensity
$\lambda_j$. The parameter vectors are $\bm{\beta}_j =
(\kappa_j,\mu_j,\nu_j,y_j(0),\chi_j,m_j)$ with each vector component being a
positive deterministic constant. We relax the condition of
inaccessibility of the origin $2 \kappa_j \mu_j > \nu_j^2$ so that we
do not limit the CDS implied volatility generated by the model.
We assume the $Z_j$'s to be standard Brownian motion processes under the
risk neutral measure. We define the following integrated quantities
which will be extensively used in the remainder of the paper

\begin{equation}
\Lambda_j(t) = \int_0^t \lambda_j(s) ds, \ \ Y_j(t) = \int_0^t y_j(s)
ds, \ \ \Psi_j(t;\bm{\beta}_j) = \int_0^t \psi_j(s;\bm{\beta}_j) ds
\label{eq:integrated_proc}
\end{equation}
In this paper, we focus on intensities without jumps, i.e. $m_j=0$.
Brigo and El-Bachir \cite{BrigoBachir08} consider in detail the tractable model with
jumps, and this extension will be applied in future work.

\subsection{Bilateral risk credit valuation adjustment for receiver CDS} \label{subs:CDSadj}
We next proceed to the valuation of the BR-CVA adjustment for the case of CDS
payoff given by Eq. (\ref{eq:cds_payoff}).  We state the
result as a Proposition.
\begin{proposition}
The $\brcva$ at time $t$ for a receiver CDS contract (protection seller) running from time
$T_a$ to time $T_b$ with premium $S$ is given by
\begin{eqnarray}
\nonumber \brcvacds_{a,b}(t,S,\TLGD{0}{1}{2}) &=& \LGD{2} \cdot \mathbb{E}_t\left\{ \mathbf{1}_{C \cup D} \cdot D(t,\tau_2) \cdot
 \left[\mathbf{1}_{\tau_1 > \tau_2} \overline{\cds}_{a,b}(\tau_2,S,
  LGD_1)\right]^+\right\} \\
\nonumber & & - \LGD{0} \cdot \mathbb{E}_t\left\{ \mathbf{1}_{A \cup B} \cdot D(t,\tau_0) \cdot
\left[- \mathbf{1}_{\tau_1 > \tau_0} \overline{\cds}_{a,b}(\tau_0,S,
 LGD_1)\right]^+\right\} \\
\label{eq:biladj}
\end{eqnarray}
\end{proposition}
\proof
We have
\begin{eqnarray}
\nonumber \mathbb{E}_t\left\{ \mathbf{1}_{C \cup D} \cdot D(t,\tau_2) \cdot
\left[ \mathrm{NPV}(\tau_2)\right]^+\right\} &=&
\nonumber \mathbb{E}_t\left\{ \mathbf{1}_{C \cup D} \cdot D(t,\tau_2) \cdot
\left[\cds_{a,b}(\tau_2,S, \LGD{1})\right]^+\right\} \\
\nonumber &=&  \mathbb{E}_t\left\{ \mathbf{1}_{C \cup D} \cdot D(t,\tau_2) \cdot
\left[\mathbf{1}_{\tau_1 > \tau_2} \overline{\cds}_{a,b}(\tau_2,S, \LGD{1})\right]^+\right\} \\
\label{eq:cds_counterparty_developed}
\end{eqnarray}
where the first equality in Eq. (\ref{eq:cds_counterparty_developed})
follows by definition, while the last equality follows from
Eq. (\ref{eq:cds_payoff_interm_develop}). Similarly, we have
\begin{eqnarray}
\nonumber \mathbb{E}_t\left\{ \mathbf{1}_{A \cup B} \cdot D(t,\tau_0) \cdot
\left[- \mathrm{NPV}(\tau_0)\right]^+\right\} &=&
\nonumber \mathbb{E}_t\left\{ \mathbf{1}_{A \cup B} \cdot D(t,\tau_0) \cdot
\left[-\cds_{a,b}(\tau_0,S, \LGD{1})\right]^+\right\} \\
\nonumber &=&  \mathbb{E}_t\left\{ \mathbf{1}_{A \cup B} \cdot D(t,\tau_0) \cdot
\left[- \mathbf{1}_{\tau_1 > \tau_0} \overline{\cds}_{a,b}(\tau_0,S, \LGD{1})\right]^+\right\} \\
\label{eq:cds_investor_developed}
\end{eqnarray}
The proof follows using the expression of BR-CVA which is given
by Eq. (\ref{eq:br-cva}).
\endproof
To summarize, in order to compute the
counterparty risk adjustment, we determine the value of the CDS contract on the reference credit ``1'' at the
point in time $\tau_2$ at which the counterparty ``2'' defaults. The
reference name ``1'' has survived this point and there is a bivariate
copula $C_{1,2}$ %given by Eq. (\ref{eq:biv_gauss_copula})
which connects the default times of the reference credit and of the counterparty ``2''. Similarly, in order to
compute the investor risk adjustment, we determine the value of the CDS contract on the reference credit ``1'' at the
point in time $\tau_0$ at which the counterparty ``0'' defaults. The
reference name ``1'' has survived this point and there is a bivariate
copula $C_{0,1}$ 
%given by the analogous of Eq. (\ref{eq:biv_gauss_copula}) 
which connects the default times of the reference credit and of the investor ``0''.
It is clear from Eq. (\ref{eq:cds_payoff_interm_develop}) and Eq. (\ref{eq:br-cva}) that the only
terms we need to know in order to compute (\ref{eq:cds_counterparty_developed}) and
 (\ref{eq:cds_investor_developed}) are

\begin{equation}
\mathbf{1}_{C \cup D} \mathbf{1}_{\tau_1 > \tau_2} \mathbb{Q}(\tau_1 > t |
\mathcal{G}_{\tau_2})
\label{eq:surv_prob_counterparty}
\end{equation}
and
\begin{equation}
 \mathbf{1}_{A \cup B} \mathbf{1}_{\tau_1 > \tau_0} \mathbb{Q}(\tau_1 > t |
\mathcal{G}_{\tau_0})
\label{eq:surv_prob_investor}
\end{equation}

In the next section, we generalize the
numerical method proposed in \cite{Brigo08} to calculate quantities
(\ref{eq:surv_prob_counterparty}) and (\ref{eq:surv_prob_investor})
for the bilateral counterparty case.

\section{Monte-Carlo Evaluation of the BR-CVA adjustment} \label{sec:numerical_algorithm}
We propose a numerical method based on Monte-Carlo simulations to
calculate the BR-CVA for the case of CDS contracts. Subsection
\ref{CIR_sim} specifies the simulation method used to generate the
sample paths of the CIR process. Subsection \ref{subs:surv_prob} gives a method
to calculate (\ref{eq:surv_prob_counterparty}), while Subsection
\ref{subs:algorithm} gives the complete numerical algorithm for calculating the
BR-CVA in (\ref{eq:biladj}). %\ref{generalprice}).
\subsection{Simulation of CIR process}\label{CIR_sim}
We use the well known fact \cite{Brigo06} that the distribution of $y(t)$ given
$y(u)$, for some $u < t$ is, up to a scale factor, a noncentral
chi-square distribution. More precisely, the transition law of $y(t)$
given $y(u)$ can be expressed as
\begin{equation}
y(t) = \frac{\nu^2 (1-e^{-\kappa(t-u)})}{4 \kappa} \chi'_d \left(\frac{4 \kappa
e^{-\kappa (t-u)}}{\nu^2(1-e^{-\kappa(t-u)})} y(u)\right)
\label{eq:y_sim}
\end{equation}
where
\begin{equation}
d = \frac{4 \kappa \mu}{\nu^2}
\end{equation}
and $\chi'_u(v)$ denotes a non-central chi-square random variable with
$u$ degrees of freedom and non centrality parameter $v$. In this way, if we know $y(0)$, we can simulate the process $y(t)$
exactly on a discrete time grid by sampling from the non-central
chi-square distribution.

\subsection{Calculation of Survival Probability}\label{subs:surv_prob}
We state the result in the form of a proposition. Such result will be
used in Subsection \ref{subs:algorithm} to develop a numerical
algorithm for computing the BR-CVA. Let us define
\begin{equation}
\overline{U}_{i,j} = 1-\exp(-\Lambda_i(\tau_j))
\end{equation}
and denote by $F_{\Lambda_i(t)}$ the cumulative distribution function of the
cumulative (shifted) intensity of the CIR process associated to name $i$,
which can be retrieved inverting the characteristic function of the integrated CIR process.
We have
\begin{proposition}
\begin{eqnarray}
\nonumber & & \mathbf{1}_{C \cup D} \mathbf{1}_{\tau_1 > \tau_2} \mathbb{Q}(\tau_1 > t |
\mathcal{G}_{\tau_2}) = \\
\nonumber & & \mathbf{1}_{\tau_2 \leq T} \mathbf{1}_{\tau_2 \leq \tau_0} \left( \mathbf{1}_{\bar A} + \mathbf{1}_{\tau_2 < t} \mathbf{1}_{\tau_1
\geq   \tau_2} \int_{\overline{U}_{1,2}}^1 F_{\Lambda_1(t)- \Lambda_1(\tau_2)}(-\log(1-u_1) - \Lambda_1(\tau_2)) dC_{1|0,2}(u_1;U_2) \right)\\
\label{eq:surv_prob_computation}
\end{eqnarray}
\text{where}
\begin{eqnarray}
\nonumber \bar{A} &=& \{t < \tau_2 < \tau_1\}\\
\nonumber C_{1|0,2}(u_1;U_2) &=& \frac{\frac{\partial C_{1,2}
(u_1,u_2)}{\partial u_2}\bigg|_{u_2 = U_2} - \frac{\partial C
(\overline{U}_{0,2},u_1,u_2)}{\partial u_2}\bigg|_{u_2 = U_2} -
\frac{\partial C_{1,2} (\overline{U}_{1,2},u_2)}{\partial
u_2}\bigg|_{u_2 = U_2} + \frac{\partial C
(\overline{U}_{0,2},\overline{U}_{1,2},u_2)}{\partial
u_2}\bigg|_{u_2 = U_2} }{1-\frac{\partial C_{0,2}
(\overline{U}_{0,2},u_2)}{\partial u_2}\bigg|_{u_2 = U_2} -
\frac{\partial C_{1,2} (\overline{U}_{1,2},u_2)}{\partial u_2}\bigg|_{u_2 = U_2} + \frac{\partial
C(\overline{U}_{0,2},\overline{U}_{1,2},u_2)}{\partial
u_2}\bigg|_{u_2 =
U_2}} \\
\label{eq:copula2first}
\end{eqnarray}
\end{proposition}
\noindent Similarly,
\begin{proposition}
\begin{eqnarray}
\nonumber & & \mathbf{1}_{A \cup B} \mathbf{1}_{\tau_1 > \tau_0} \mathbb{Q}(\tau_1 > t |
\mathcal{G}_{\tau_0}) = \\
\nonumber & & \mathbf{1}_{\tau_0 \leq T} \mathbf{1}_{\tau_0 \leq \tau_2} \left(\mathbf{1}_{\bar B} + \mathbf{1}_{\tau_0 < t} \mathbf{1}_{\tau_1
\geq \tau_0} \int_{\overline{U}_{1,0}}^1 F_{\Lambda_1(t)- \Lambda_1(\tau_0)}(-\log(1-u_1) - \Lambda_1(\tau_0) ) dC_{1|2,0}(u_1;U_0)\right) \\
\label{eq:surv_prob_computation_sec0}
\end{eqnarray}
\text{where}
\begin{eqnarray}
\nonumber \bar{B} &=& \{t < \tau_0 < \tau_1\}\\
\nonumber C_{1|2,0}(u_1;U_0) &=& \frac{\frac{\partial C_{0,1} (u_0,u_1)}{\partial
u_0}\bigg|_{u_0 = U_0} - \frac{\partial C
(u_0,u_1,\overline{U}_{2,0})}{\partial u_0}\bigg|_{u_0 = U_0} -
\frac{\partial C_{0,1} (u_0,\overline{U}_{1,0})}{\partial
u_0}\bigg|_{u_0 = U_0} + \frac{\partial C
(u_0,\overline{U}_{1,0},\overline{U}_{2,0})}{\partial
u_0}\bigg|_{u_0 = U_0}}{1-\frac{\partial C_{0,2}
(u_0,\overline{U}_{2,0})}{\partial u_0}\bigg|_{u_0 = U_0} -
\frac{\partial C_{0,1}
(u_0,\overline{U}_{1,0})}{\partial u_0}\bigg|_{u_0 = U_0} + \frac{\partial C
(u_0,\overline{U}_{1,0},\overline{U}_{2,0})}{\partial
u_0}\bigg|_{u_0 = U_0}} \\
\label{eq:copula0first}
\end{eqnarray}
\end{proposition}
The proofs of the propositions are reported in Appendix \ref{sec:surv_prob_proof}.

\subsection{The numerical BR-CVA adjustment algorithm}\label{subs:algorithm}
We give the pseudo code of the numerical algorithm used to calculate the BR-CVA
 for the CDS payer and receiver. In the pseudo-code
below, the variable $\alpha_i$ represents the time elapsing between
payment period $t_{i-1}$ and $t_i$ measured in years, the variable
$\Delta$ represents the fineness of the grid used to evaluate the
 integral of the survival probability in
Eq. (\ref{eq:surv_prob_computation}) and the variable
$\delta$ represents the fineness of the time grid used to evaluate the
integral in Eq. (\ref{eq:cds_payoff_interm_develop}). The variable $x_{\max}$
represents the maximum $x$ value for which the cumulative distribution
 function is implied from the characteristic function. The expression
CDF($x_k$) corresponds to a subroutine call which calculates
the cumulative distribution function of the integrated CIR process at $x_k$. This can be done by
 inversion of the characteristic function of the integrated CIR
 process using Fourier transform methods such as in \cite{Chou05}. The inputs to the main procedure
 \CalculateAdjustment are the number $N$ of Monte-Carlo runs and the
market quote $S_1$ of the 5 year CDS spread of the reference entity.

%============= Pseudo Code of the algorithm =============================%
\begin{algorithm}[h!]
\caption{$[\overline{\text{BR-CVA\_R}}, \overline{\text{BR-CVA\_P}} ]$ =
  \CalculateAdjustment(N, S)}
\floatname{algorithm}{Procedure}
\label{alg:Calculate_Adjustment}
\begin{algorithmic}
\FOR{$i = 1:N$}
\STATE Generate $\tau_0$, $\tau_1$, and $\tau_2$ using
Eq. (\ref{eq:unif_implied}) and Eq. (\ref{eq:trivar_copula}).
\IF{$\tau_2 < \tau_0$ and $\tau_2 < T_b$}
\IF{$\tau_1 > \tau_2$}
\STATE $[$BR-CVA\_R\_2, BR-CVA\_P\_2$]$ = CDSAdjust($T_{\gamma(\tau_2)}$, $S$, $\LGD{1}$, 2)
\STATE CUM\_BR-CVA\_R =  CUM\_BR-CVA\_R + $\LGD{2} \cdot$ BR-CVA\_R\_2
\STATE CUM\_BR-CVA\_P =  CUM\_BR-CVA\_P + $\LGD{2} \cdot$ BR-CVA\_P\_2
\ENDIF
\ENDIF
\IF{$\tau_0 < \tau_2$ and $\tau_0 < T_b$}
\IF{$\tau_1 > \tau_0$}
\STATE $[$BR-CVA\_R\_0, BR-CVA\_P\_0$]$ = CDSAdjust($T_{\gamma(\tau_0)}$, $S$, $\LGD{1}$, 0)
\STATE CUM\_BR-CVA\_R = CUM\_BR-CVA\_R - $\LGD{0} \cdot$ BR-CVA\_R\_0
\STATE CUM\_BR-CVA\_P = CUM\_BR-CVA\_P - $\LGD{0} \cdot$ BR-CVA\_P\_0
\ENDIF
\ENDIF
\ENDFOR
\STATE $\overline{\text{BR-CVA\_R}}$ = CUM\_BR-CVA\_R / $N$
\STATE $\overline{\text{BR-CVA\_P}}$ = CUM\_BR-CVA\_P / $N$
\end{algorithmic}
\end{algorithm}

\begin{algorithm}[h!]
\caption{[$CDS_R, CDS_P$] = \CDSAdjust($T_j ,S$, $\LGD{1}$, $index$)}
\floatname{algorithm}{Procedure}
\label{alg:Calculate_CDS}
\begin{algorithmic}
\STATE $Term_1 = Term_2 = Term_3 = 0$
\STATE $t_{start} = \max(T_a, T_j)$
\STATE $Q_{prev} = $ ComputeProb($t_{start}$, $T_j$, $index$)
\FOR{$t = t_{start}+\delta:\delta:T_b$}
\STATE $Q_{curr} = $ ComputeProb($t, T_j, index$)
\STATE $Term_1 = Term_1 + D(T_j,t - \delta) (t - \delta -
T_{\gamma(t-\delta)-1}) (Q_{curr} - Q_{prev})$
\STATE $Term_3 = Term_3 + D(T_j, t-\delta) (Q_{curr} - Q_{prev})$
\STATE $Q_{prev} = Q_{curr}$
\ENDFOR
\FOR{$t_i = t_{start} + \alpha_i: \alpha_i: T_b$}
\STATE $Q_{curr} = $ ComputeProb($t_i, T_j, index$)
\STATE $Term_2 = Term_2 + \alpha_i \cdot D(T_j, t_i) \cdot Q_{curr}$
\ENDFOR
\STATE $CDS_{val} = S \cdot (Term_2 - Term_1)$ + $\LGD{1} \cdot Term_3$
\IF{$index == 2$}
\STATE $CDS_R = D(t, T_j) \cdot \max(CDS_{val},0)$
\STATE $CDS_P = D(t, T_j) \cdot \max(-CDS_{val},0)$
\ENDIF
\IF{$index == 0$}
\STATE $CDS_R = D(t, T_j) \cdot \max(-CDS_{val},0)$
\STATE $CDS_P = D(t, T_j) \cdot \max(CDS_{val},0)$
\ENDIF
\end{algorithmic}
\end{algorithm}

\begin{algorithm}[h!]
%\algsetup{indent=2em}
%\algsetup{
%linenosize=\small,
%linenodelimiter=.}
\caption{$Q_i$ = \ComputeProb($t, T_j, index$)}
\floatname{algorithm}{Procedure}
\label{alg:Compute_Prob}
\begin{algorithmic}
\STATE $U_{index} = 1 - \exp\{-Y_{index}(T_j) - \Psi_{index}(T_j;\bm{\beta}_{index}) \}$
\STATE $\bar{U}_1 = 1 - \exp\{-Y_1(T_j) - \Psi_1(T_j;\bm{\beta}_1)\}$
\FOR{$x_k = 0:\Delta:x_{\max}$}
\STATE $p_k$ = CDF($x_k$)
\STATE $u_k = 1 - \exp\{-x_k - \Psi_1(t)\}$
\IF {index == 2}
\STATE Compute $f_k = C_{1|0,2}(u_k;U_2)$ using Eq. (\ref{eq:copula2first})
\ELSE
\STATE Compute $f_k = C_{1|2,0}(u_k;U_0)$ using Eq. (\ref{eq:copula0first})
\ENDIF
\ENDFOR
\STATE $Q_i = \sum_{(u_k,p_k,f_k): u_k > \bar{U}_1} p_k (f_{k+1} - f_k)$
\end{algorithmic}
\end{algorithm}

\section{Numerical Results} \label{sec:case_study}
We consider an investor (name ``0'') trading a five-years CDS contract on a
reference name (name ``1'') with a counterparty (name ``2''). Both the
investor and the counterparty are subject to default risk. We experiment on
different levels of credit risk and credit risk volatility of the
three names, which are specified by the parameters of the
CIR processes in Table \ref{tab:CIR_param}.
\begin{table}[h!]
\begin{center}
\begin{tabular}{|c|c|c|c||c|c|}
  \hline
  Credit Risk Levels & $y(0)$ & $\kappa$ & $\mu$ & Credit Risk
  volatilities & $\nu$ \\
  \hline
  \hline
  low  & 0.00001 & 0.9 & 0.0001 & low & 0.01 \\
  \hline
  middle & 0.01 & 0.80 & 0.02 & middle & 0.2 \\
  \hline
  high  & 0.03 & 0.50 & 0.05 & high & 0.5 \\
  \hline
\end{tabular}
\caption{The credit risk levels and credit risk volatilities parameterizing the CIR processes}
\label{tab:CIR_param}
\end{center}
\end{table}
We recall that the survival probabilities associated with a CIR
intensity process are given by
\begin{eqnarray}
\nonumber \mathbb{Q}(\tau_i > t) &:=& \mathbb{E}[e^{-Y_i(t)}] \\
&=& P^{CIR}(0,t,\bm{\beta}_i)
\label{eq:surv_cir}
\end{eqnarray}
where $P^{CIR}(0,t,\bm{\beta}_i)$ is the price at time 0 of a zero coupon bond
 maturing at time $t$ under a stochastic interest rate dynamics given by the CIR process
  \cite{Brigo06}, with $\bm{\beta}_i = (y_i(0), \kappa_i, \mu_i, \nu_i)$ being
  the vector of CIR parameters, $i = 0, 1, 2$.

We report in Table \ref{tab:spreads} the break-even spreads zeroing
  (\ref{eq:cds_payoff}) in $S$, with survival probabilities
  given by Eq. (\ref{eq:surv_cir}) and CIR parameters $\bm{\beta}_{low}$,
 $\bm{\beta}_{middle}$, $\bm{\beta}_{high}$ obtained
 from Table \ref{tab:CIR_param}.

\begin{table}
\begin{center}
\begin{tabular}{|c|c|c|c|}
\hline
Maturity & Low Risk & Middle Risk & High risk
\\ \hline
\hline
1y & 0&92&234\\\hline
2y & 0&104&244\\\hline
3y & 0&112&248\\\hline
4y & 1&117&250\\\hline
5y & 1&120&251\\\hline
6y & 1&122&252\\\hline
7y & 1&124&253\\\hline
8y & 1&125&253\\\hline
9y & 1&126&254\\\hline
10y & 1&127&254\\\hline
\end{tabular}
\caption{Break-even spreads in basis points generated using the
  parameters of the CIR processes in Table \ref{tab:CIR_param}.
The first column is generated using low credit
  risk and credit risk volatility. The second column is generated
  using middle credit risk and credit risk volatility. The third column
  is generated using high credit risk and credit risk volatility.}
\label{tab:spreads}
\end{center}
\end{table}
The evaluation time $t$ and the starting time $T_a$ of the CDS contract are
both set to zero. The end time $T_b$ of the contract is set to five
years. It is assumed that payments are exchanged every three months.
The loss given defaults of the low, middle and high risk entity are
respectively set to $\LGD{low} = 0.6$, $\LGD{middle} = 0.65$,
$\LGD{high}=0.7$.  We assume that the spreads in Table
\ref{tab:spreads} are the spreads quoted in the markets for the three
names under consideration. We recover the integrated shift $\Psi(t;\bm{\beta})$
which makes the model survival probabilities consistent with the
market survival probabilities coming from Table \ref{tab:spreads}
whenever we change the CIR parameters. In mathematical terms, for any
$i=0,1,2$, we impose that
\begin{eqnarray}
\nonumber \mathbb{Q}(\tau_i > t)_{model} &:=& \mathbb{E}[e^{-\Lambda_i(t)}] \\
 &=& \mathbb{Q}(\tau_i > t)_{market}
\label{eq:consistent_prob}
\end{eqnarray}
The market survival probability for name ``$i$'' is bootstrapped from
the market CDS quotes reported in Table \ref{tab:spreads}. Such
bootstrap procedure is performed assuming a piecewise linear hazard
rate function. From the definition of the integrated process
$\Psi_j(t;\bm{\beta}_j)$ given in Eq. (\ref{eq:integrated_proc}), we
can restate Eq. (\ref{eq:consistent_prob}) as
\begin{eqnarray}
\nonumber \Psi_i(t;\bm{\beta}_i) &=&   \log \left( \frac{\mathbb{E}[e^{-Y_i(t)}]}{\mathbb{Q}(\tau_i >t)_{market}} \right) \\
&=& \log \left(\frac{P^{CIR}(0,t,\bm{\beta}_i)}{\mathbb{Q}(\tau_i >t)_{market}}\right)
\label{eq:shift_integrated}
\end{eqnarray}
where the last equality in Eq. (\ref{eq:shift_integrated}) follows
from Eq. (\ref{eq:surv_cir}).
We first study a case where the quoted market spreads of the investor has
a low credit risk profile ($\beta_{low}$), the reference entity has high
 credit risk profile ($\beta_{high}$), while the counterparty has middle
 credit risk ($\beta_{middle}$). We vary the correlation
 between reference credit and counterparty as well as the credit risk
 volatility $\nu_1$ of the reference credit. Since the focus is mostly
 on credit spreads volatility, we calculate the implied CDS volatility
 produced by the choice of parameters
$\bm{\beta}_1 = (y_0(1), \mu_1, \kappa_1, \nu_1)$ for a
hypothetical CDS option, maturing in one year and in case of
exercise entering into a four year CDS contract on the underlying
reference credit ``1''. The objective of the experiments is to measure
the impact of correlation and credit spreads volatility on the BR-CVA.
The triple $(x,y,z)$ represents the   correlation of the trivariate Gaussian copula, with $x$ denoting the correlation between the
  investor and reference credit, $y$ denoting the correlation between the investor and the counterparty
 and $z$ denoting the correlation between the reference credit and the
  counterparty. The values $\overline{\text{BR-CVA\_P}}$ and $\overline{\text{BR-CVA\_R}}$ are
respectively the Monte-Carlo estimates of the CDS payer and receiver
counterparty risk adjustments. The theoretical formula for payer risk
adjustment is given by
\begin{eqnarray}
\nonumber \mathrm{\LGD{2} \cdot \mathbb{E}_t\left\{ \mathbf{1}_{C \cup D} \cdot D(t,\tau_2) \cdot \left[-\mathrm{NPV}(\tau_2)\right]^+\right\}} \\
  - \mathrm{\LGD{0} \cdot \mathbb{E}_t\left\{ \mathbf{1}_{A \cup B}  \cdot D(t,\tau_0) \cdot \left[\mathrm{NPV}(\tau_0)\right]^+\right\}}
\end{eqnarray}
which follows from adapting the formula given in
Eq. (\ref{eq:biladj}), given for the receiver counterparty
adjustment, to the case of the payer counterparty adjustment.
Tables \ref{tab:results_experim} and \ref{tab:results_experim_2}
report the results obtained.

\begin{table}
\begin{small}\begin{tabular}{|l|l|c|c|c|c|c|c|}
\hline
\textbf{$(r_{01},r_{02},r_{12})$}&\textbf{Vol. parameter $\nu_1$}&\textbf{0.01}&\textbf{0.10}&\textbf{0.20}&\textbf{0.30}&\textbf{0.40}&\textbf{0.50}\\
& \textbf{CDS Impled vol}&\textbf{1.5\%}&\textbf{15\%}&\textbf{28\%}&\textbf{37\%}&\textbf{42\%}&\textbf{42\%}\\
 \hline
\textbf{(0, 0, -0.99)}&$\overline{\text{BR-CVA\_P}}$ &0.0(0.0)&0.0(0.0)&0.0(0.0)&0.0(0.0)&-0.0(0.0)&-0.0(0.0)\\
&$\overline{\text{BR-CVA\_R}}$&29.3(1.5)&29.7(1.5)&29.8(1.5)&29.8(1.5)&29.8(1.5)&29.6(1.5)\\
\hline
\textbf{(0, 0, -0.90)}&$\overline{\text{BR-CVA\_P}}$&0.0(0.0)&0.0(0.0)&0.0(0.0)&0.0(0.0)&-0.0(0.0)&-0.0(0.0)\\
&$\overline{\text{BR-CVA\_R}}$&29.6(1.5)&29.2(1.5)&29.1(1.5)&29.3(1.5)&29.6(1.5)&29.0(1.5)\\
\hline
\textbf{(0, 0, -0.60)}&$\overline{\text{BR-CVA\_P}}$&0.0(0.0)&0.0(0.0)&0.0(0.0)&0.1(0.0)&0.4(0.2)&0.6(0.2)\\
&$\overline{\text{BR-CVA\_R}}$&27.0(1.4)&27.2(1.4)&26.4(1.4)&26.6(1.4)&25.4(1.3)&25.2(1.3)\\
\hline
\textbf{(0, 0, -0.20)}&$\overline{\text{BR-CVA\_P}}$&0.0(0.0)&0.4(0.1)&1.3(0.2)&2.0(0.3)&2.6(0.4)&3.2(0.5)\\
&$\overline{\text{BR-CVA\_R}}$&10.3(0.7)&10.4(0.7)&10.9(0.7)&12.1(0.7)&12.8(0.8)&13.0(0.8)\\
\hline
\textbf{(0, 0, 0)}&$\overline{\text{BR-CVA\_P}}$&4.8(0.3)&5.1(0.4)&5.4(0.5)&7.4(0.8)&5.9(0.6)&5.6(0.7)\\
&$\overline{\text{BR-CVA\_R}}$&0.0(0.0)&0.5(0.1)&2.1(0.2)&3.9(0.3)&5.0(0.3)&6.3(0.4)\\
\hline
\textbf{(0, 0, 0.20)}&$\overline{\text{BR-CVA\_P}}$&25.9(1.5)&25.1(1.5)&23.2(1.5)&21.1(1.6)&15.9(1.3)&12.8(1.2)\\
&$\overline{\text{BR-CVA\_R}}$&0.0(0.0)&0.0(0.0)&0.2(0.0)&0.5(0.1)&0.7(0.1)&1.1(0.1)\\
\hline
\textbf{(0, 0, 0.60)}&$\overline{\text{BR-CVA\_P}}$&72.0(4.9)&72.0(4.8)&65.2(4.4)&57.1(4.0)&50.8(3.6)&40.4(3.0)\\
&$\overline{\text{BR-CVA\_R}}$&0.0(0.0)&0.0(0.0)&0.0(0.0)&0.0(0.0)&0.0(0.0)&0.1(0.0)\\
\hline
\textbf{(0, 0, 0.90)}&$\overline{\text{BR-CVA\_P}}$&68.4(6.1)&73.8(6.3)&69.2(5.8)&65.2(5.4)&61.5(5.0)&62.2(4.9)\\
&$\overline{\text{BR-CVA\_R}}$&0.0(0.0)&0.0(0.0)&0.0(0.0)&0.0(0.0)&0.0(0.0)&0.0(0.0)\\
\hline
\textbf{(0, 0, 0.99)}&$\overline{\text{BR-CVA\_P}}$&13.2(2.8)&28.4(4.0)&39.4(4.4)&51.9(4.9)&54.8(5.0)&67.6(5.4)\\
&$\overline{\text{BR-CVA\_R}}$&0.0(0.0)&0.0(0.0)&0.1(0.0)&0.1(0.0)&0.1(0.0)&0.4(0.3)\\
\hline
\end{tabular}
\end{small}
\caption{BR-CVA in basis points for the case when $y_0(0) = 0.0001, \kappa_0 = 0.01, \mu_0=0.001, \nu_0 = 0.01$,
 $y_0(2) = 0.01, \kappa_2 = 0.8, \mu_2=0.02, \nu_2 = 0.2$. The name ``1'' has CIR parameters
 $y_0(1) = 0.03, \kappa_1 = 0.5, \mu_1=0.05$,  while the credit spreads volatility $\nu_1$ is varied across scenarios.
 The numbers within round brackets represent the Monte-Carlo standard error. The CDS contract on the reference credit has a five-years maturity.}
\label{tab:results_experim}
\end{table}

\begin{table}
\begin{small}\begin{tabular}{|l|l|c|c|c|c|c|c|}
\hline
\textbf{$(r_{01},r_{02},r_{12})$}&\textbf{Vol. parameter $\nu_1$}&\textbf{0.01}&\textbf{0.10}&\textbf{0.20}&\textbf{0.30}&\textbf{0.40}&\textbf{0.50}\\
& \textbf{CDS Impled vol}&\textbf{1.5\%}&\textbf{15\%}&\textbf{28\%}&\textbf{37\%}&\textbf{42\%}&\textbf{42\%}\\
 \hline
\textbf{(0, 0, -0.99)}&$\overline{\text{BR-CVA\_P}}$&0.0(0.0)&0.0(0.0)&0.0(0.0)&0.0(0.0)&0.0(0.0)&0.0(0.0)\\
&$\overline{\text{BR-CVA\_R}}$&28.8(1.4)&29.2(1.4)&28.3(1.4)&28.0(1.4)&29.3(1.4)&28.9(1.4)\\
\hline
\textbf{(0, 0, -0.90)}&$\overline{\text{BR-CVA\_P}}$&0.0(0.0)&0.0(0.0)&0.0(0.0)&0.0(0.0)&0.1(0.0)&0.1(0.0)\\
&$\overline{\text{BR-CVA\_R}}$&28.9(1.4)&29.0(1.4)&28.2(1.4)&28.7(1.4)&28.8(1.4)&28.9(1.4)\\
\hline
\textbf{(0, 0, -0.60)}&$\overline{\text{BR-CVA\_P}}$&0.0(0.0)&0.0(0.0)&0.1(0.1)&0.1(0.0)&0.8(0.2)&0.2(0.1)\\
&$\overline{\text{BR-CVA\_R}}$&26.8(1.3)&26.5(1.3)&25.7(1.3)&25.1(1.3)&24.9(1.3)&25.0(1.3)\\
\hline
\textbf{(0, 0, -0.20)}&$\overline{\text{BR-CVA\_P}}$&0.0(0.0)&0.4(0.1)&1.3(0.2)&2.4(0.4)&2.8(0.5)&2.0(0.4)\\
&$\overline{\text{BR-CVA\_R}}$&9.7(0.6)&9.7(0.6)&10.4(0.7)&11.7(0.7)&12.7(0.7)&13.0(0.7)\\
\hline
\textbf{(0, 0, 0)}&$\overline{\text{BR-CVA\_P}}$&4.8(0.2)&5.3(0.4)&6.0(0.5)&6.6(0.7)&5.2(0.7)&5.1(0.7)\\
&$\overline{\text{BR-CVA\_R}}$&0.0(0.0)&0.5(0.1)&2.1(0.2)&3.8(0.3)&5.0(0.3)&6.0(0.3)\\
\hline
\textbf{(0, 0, 0.20)}&$\overline{\text{BR-CVA\_P}}$&26.6(1.5)&26.1(1.5)&23.6(1.5)&19.7(1.4)&16.7(1.4)&13.8(1.4)\\
&$\overline{\text{BR-CVA\_R}}$&0.0(0.0)&0.0(0.0)&0.1(0.0)&0.4(0.0)&0.7(0.1)&1.0(0.1)\\
\hline
\textbf{(0, 0, 0.60)}&$\overline{\text{BR-CVA\_P}}$&76.4(4.8)&74.1(4.6)&68.1(4.4)&60.8(4.0)&52.2(3.7)&42.3(3.1)\\
&$\overline{\text{BR-CVA\_R}}$&0.0(0.0)&0.0(0.0)&0.0(0.0)&0.0(0.0)&0.1(0.0)&0.1(0.0)\\
\hline
\textbf{(0, 0, 0.90)}&$\overline{\text{BR-CVA\_P}}$&75.3(6.1)&76.1(6.1)&74.4(5.9)&68.8(5.5)&62.2(5.2)&64.4(4.9)\\
&$\overline{\text{BR-CVA\_R}}$&0.0(0.0)&0.0(0.0)&0.0(0.0)&0.0(0.0)&0.0(0.0)&0.0(0.0)\\
\hline
\textbf{(0, 0, 0.99)}&$\overline{\text{BR-CVA\_P}}$&12.6(2.6)&24.9(3.6)&41.3(4.5)&51.8(4.9)&55.3(4.9)&65.9(5.3)\\
&$\overline{\text{BR-CVA\_R}}$&0.0(0.0)&0.0(0.0)&0.1(0.0)&0.1(0.0)&0.1(0.0)&0.1(0.0)\\
\hline
\end{tabular}
\end{small}
\caption{BR-CVA in basis points for the case when $y_0(0) = 0.0001, \kappa_0 = 0.01, \mu_0=0.001, \nu_0 = 0.01$,
 $y_0(2) = 0.01, \kappa_2 = 0.8, \mu_2=0.02, \nu_2 = 0.01$. The name ``1'' has CIR parameters
 $y_0(1) = 0.03, \kappa_1 = 0.5, \mu_1=0.05$,  while the credit spreads volatility $\nu_1$ is varied across scenarios.
 The numbers within round brackets represent the Monte-Carlo standard error. The CDS contract on the reference credit has a five-years maturity.}
\label{tab:results_experim_2}
\end{table}
The scenarios considered in Table \ref{tab:results_experim} and
\ref{tab:results_experim_2} assume an investor with an extremely low
credit risk profile and thus are similar to those considered in
\cite{Brigo08}, where the investor is assumed to be
default-free. Therefore, it is not surprising that we find similar
results. Similarly to them, we can see that the BR-CVA for the payer investor tends to be monotonically increasing with
the correlation $r_{12}$, whereas the BR-CVA for the receiver investor appears to be monotonically decreasing.
This is expected since for low and negative correlation values, defaults of the counterparty come with reductions in default risk of the reference entity, thus the receiver investor holds an option which is in the money, but at the counterparty default he only gets a fraction of it proportional to the recovery value of the counterparty. On the contrary, for high and positive correlation values, defaults of the counterparty come with increases in default risk of the reference entity, thus the payer investor holds an option which is in the money, but at the counterparty default he only gets a fraction of it proportional to the recovery value of the counterparty.
While these patterns are intuitive, there is a particular pattern in correlation that is seemingly counterintuitive.

\begin{figure}
\vspace{-5.1cm}
\begin{center}
\hspace{-0.8cm}
\includegraphics[scale=0.6]{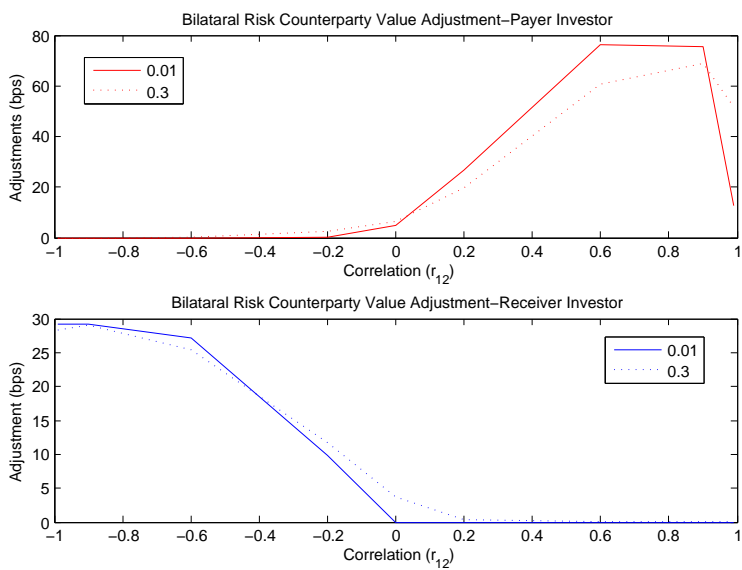}
\vspace{-4.2cm}
\caption{Patterns of the $\overline{\text{BR-CVA}}$ of the CDS contract for payer and receiver investor and for low (0.1) and high (0.3) reference credit spreads volatility $\nu_1$. We have that $\nu_2 = 0.01$ and $\nu_0 = 0.01$ }
\label{fig:brcva001}
\end{center}
\end{figure}

\begin{remark}{\bf (Decreasing wrong way risk with low credit spread volatility and high correlation).}
In general, we expect the BR-CVA for the payer investor to increase with correlation between the underlying reference credit and the counterparty default. However, for correlation increasing beyond relatively large values, the BR-CVA for the payer investor goes down instead, if the credit spreads volatility of the reference entity is small.
Consider, for example, the case when the reference entity and the counterparty
are $99\%$ correlated, so that the exponential triggers
$\xi_1$ and $\xi_2$ are almost identical. If we consider the
scenario where $\nu_1 = \nu_2 = 0.01$, then the intensity of both
processes are almost deterministic, with $\lambda_1 > \lambda_2$
having the name  ``1'' high credit risk and name ``2'' middle credit
risk. Thus, to a first approximation $\tau_1 =
\frac{\xi_1}{\lambda_1}$ is higher than $\tau_2 =
\frac{\xi_2}{\lambda_2}$, and consequently the reference credit
always defaults before the counterparty, resulting in no adjustment.
Table \ref{tab:results_experim_2} has qualitatively a similar
behavior. This is visible also in the right side of the graphs in the upper Figure~\ref{fig:brcva001}, where the pattern goes down at the very end for correlations beyond 0.9.

This seemingly counterintuitive feature is improved if credit spread volatility becomes large (and in line with realistic CDS implied volatilities, see Brigo (2005)\cite{BrigoCDSMM}). This is evident from the last column of Table \ref{tab:results_experim_2}, and is due to the increased randomness in the default times, that now can cross each other in more scenarios.
\end{remark}

%although the larger value of the counterparty volatility
%$\nu_2 = 0.2$ tends to smooth out the negligible BR-CVA obtained
%in Table \ref{tab:results_experim} for small values of
%reference credit volatility and high default correlation between counterparty
%and reference credit.
To further illustrate the transition from large (small) to small (large) adjustments for the receiver (payer) investor
and illustrate the effect of correlation and credit spreads volatility, we display in Figure \ref{fig:brcva001}
the behavior of the bilateral risk adjustment for two different values of credit spreads volatility.

Tables \ref{tab:one_cor} and \ref{tab:three_cor}
report the BR-CVA adjustments for the Payer CDS under a set of five
different riskiness scenarios. The assignments of
credit risks in Table \ref{tab:CIR_param} to the three names determine
the scenario in place.
We have:
\begin{itemize}
\item Scenario 1 (\emph{Base Scenario}). The investor has low credit
  risk, the reference entity has high credit risk, and the counterparty
  has middle credit risk
\item Scenario 2 (\emph{Risky counterparty}). The investor has low
  credit risk, the reference entity has middle credit risk, and the counterparty has high credit risk.
\item  Scenario 3 (\emph{Risky investor}). The counterparty has low
  credit risk, the reference entity has middle credit risk, and the
  investor has high credit risk.
\item Scenario 4 (\textit{Risky Ref}). Both investor and counterparty
  have middle credit risk, while the reference entity has high credit risk.
\item Scenario 5 (\emph{Safe Ref}). Both investor and counterparty
  have high credit risk, while the reference entity has low credit risk.
\end{itemize}

\begin{table}
\begin{center}
\begin{small}\begin{tabular}{|l|c|c|c|c|c|}
\hline
\textbf{$(r_{01},r_{02},r_{12})$} &\textbf{Base Scenario} &\textbf{Risky Counterparty} &\textbf{Risky Investor} &\textbf{Risky Ref} &\textbf{Safe Ref} \\
\hline
\textbf{(0, 0, 0)}&6.0(0.4) &3.6(0.2) &-0.8(0.0) &-0.0(0.0) &5.5(0.4) \\
\hline
\textbf{(0, 0, 0.1)}&15.1(0.8) &12.6(0.5) &-0.8(0.0) &-0.0(0.0) &13.2(1.0) \\
\hline
\textbf{(0, 0, 0.3)}&37.0(2.0) &37.4(1.5) &-0.8(0.1) &0.1(0.0) &34.6(2.1) \\
\hline
\textbf{(0, 0, 0.6)}&73.7(4.4) &92.5(3.8) &-0.3(0.5) &0.7(0.1) &67.3(4.3) \\
\hline
\textbf{(0, 0, 0.9)}&83.4(6.0) &207.8(8.2) &-0.8(0.0) &1.5(0.3) &75.7(5.8) \\
\hline
\textbf{(0, 0, 0.99)}&26.0(3.5) &316.9(12.5) &-0.8(0.0) &1.8(0.5) &22.6(3.4) \\
\hline
\textbf{(0, 0.1, 0)}&6.0(0.4) &3.6(0.2) &-0.8(0.0) &-0.0(0.0) &4.2(0.3) \\
\hline
\textbf{(0, 0.3, 0)}&5.8(0.4) &3.6(0.2) &-0.8(0.0) &-0.0(0.0) &3.7(0.3) \\
\hline
\textbf{(0, 0.6, 0)}&6.0(0.4) &3.6(0.2) &-0.8(0.1) &-0.0(0.0) &5.9(0.4) \\
\hline
\textbf{(0, 0.9, 0)}&0.6(4.5) &3.5(0.2) &-0.8(0.0) &0.0(0.0) &15.0(1.5) \\
\hline
\textbf{(0.1, 0, 0)}&6.0(0.4) &3.6(0.2) &-0.1(0.0) &-0.0(0.0) &6.5(0.4) \\
\hline
\textbf{(0.3, 0, 0)}&6.0(0.4) &3.6(0.2) &0.0(0.0) &0.0(0.0) &6.6(0.4) \\
\hline
\textbf{(0.6, 0, 0)}&6.0(0.4) &3.6(0.2) &0.0(0.0) &0.0(0.0) &6.5(0.4) \\
\hline
\textbf{(0.9, 0, 0)}&5.9(0.4) &3.6(0.2) &0.0(0.0) &-0.1(0.0) &6.6(0.4) \\
\hline
\textbf{(0.99, 0, 0)}&5.9(0.4) &3.6(0.2) &-1.8(0.1) &-0.1(0.0) &6.7(0.4) \\
\hline
\end{tabular}
\end{small}
\end{center}
\caption{BR-CVA under five different riskiness scenarios. The CIR volatilities
  are set to $\nu_0 =  \nu_1 =  \nu_2 = 0.1$. The correlation triple
has two non-zero entries}
\label{tab:one_cor}
\end{table}

\begin{table}
\begin{center}
\begin{small}
\begin{tabular}{|l|c|c|c|c|c|}
\hline
\textbf{$(r_{01}, r_{02}, r_{20})$} &\textbf{Base Scenario} &\textbf{Risky Counterparty} &\textbf{Risky Investor} &\textbf{Risky Ref} &\textbf{Safe Ref} \\
\hline
\textbf{(0.7, 0.4, 0.3)}&37.0(2.1) &36.4(1.4) &-0.0(0.0) &-0.0(0.0) &8.4(0.7) \\
\hline
\textbf{(0.7, 0.3, 0.4)}&49.0(2.8) &51.4(2.1) &0.1(0.1) &0.2(0.0) &25.8(1.7) \\
\hline
\textbf{(0.4, 0.7, 0.3)}&35.3(2.0) &35.5(1.4) &0.1(0.1) &0.1(0.0) &27.0(1.7) \\
\hline
\textbf{(0.4, 0.3, 0.7)}&79.8(5.0) &117.4(4.9) &0.6(0.5) &0.6(0.1) &74.0(4.9) \\
\hline
\textbf{(0.3, 0.4, 0.7)}&80.8(5.1) &117.8(4.9) &0.0(0.0) &0.6(0.1) &73.6(5.5) \\
\hline
\textbf{(0.3, 0.7, 0.4)}&45.8(2.7) &50.0(2.0) &0.0(0.0) &0.4(0.0) &50.9(3.3) \\
\hline
\textbf{(0.3, 0.3, 0.3)}&36.4(2.0) &36.7(1.4) &0.0(0.0) &0.1(0.0) &28.2(1.7) \\
\hline
\textbf{(0.4, 0.4, 0.4)}&47.9(2.8) &50.8(2.1) &0.0(0.0) &0.2(0.0) &39.0(2.4) \\
\hline
\textbf{(0.7, 0.7, 0.7)}&77.4(5.0) &113.6(4.8) &0.0(0.0) &0.6(0.1) &82.5(6.5) \\
\hline
\end{tabular}
\end{small}
\end{center}
\caption{BR-CVA under five different riskiness scenarios. The CIR volatilities
are set to $\nu_0 =  \nu_1 =  \nu_2 = 0.1$. The correlation triple has all non-zero entries}
\label{tab:three_cor}
\end{table}
Tables \ref{tab:one_cor} and \ref{tab:three_cor} clearly show the effect of the
wrong way risk. For example, looking at the second column, we can see that as the
correlation between counterparty and reference entity gets larger,
the BR-CVA increases significantly. This is because (1) the counterparty is the riskiest name and
(2) the high positive correlation makes the spread of the reference entity larger at the counterparty
default, thus the option on the residual NPV for the payer investor will be deep into the money and worth
more. If the only correlation is between reference credit
and counterparty, then the adjustment is the largest, while when the counterparty is less correlated with the reference entity,
but also positively correlated with the investor (see second column of Table \ref{tab:three_cor}), then the adjustment tends to
decrease as the number of scenarios where the counterparty is the earliest to default is smaller.

The first six rows of Table \ref{tab:one_cor} show how much
the BR-CVA adjustment is driven by correlation and credit spreads volatility.
When the counterparty and reference entity are loosely correlated, then the adjustments
in Scenario 1 and Scenario 2 are very similar, although the counterparty is riskier in Scenario 2.
However, as the correlation increases, the BR-CVA adjustment in Scenario 2 becomes significantly larger than
the corresponding adjustment in Scenario 1. Differently from Tables
\ref{tab:results_experim} and \ref{tab:results_experim_2}, however, we have
that low credit risk volatility values has the opposite effect and amplify
the adjustment; this is because for 99\% correlation between reference
credit and counterparty, we now have that the counterparty always defaults
earlier being riskier.

Consistently with the results in Table \ref{tab:results_experim} and
\ref{tab:results_experim_2}, no BR-CVA takes place in scenario
4. This is because the reference entity has the
highest credit risk profile and since the credit risk volatilities of the three
names are relatively low, the reference entity always defaults first, thus resulting in no adjustment taking place.

\section{Application to a market scenario}\label{sec:LehmanExperiment}
We apply the methodology to calculate the mark-to-market price of a
five-year CDS contract between British Airways (counterparty) and
Lehman Brothers (investor) on the default of Royal Dutch Shell
(reference credit). We consider two CDS contracts. In the first
contract Lehman Brothers buys 5-year protection on Shell from British Airways
on January 5, 2006. In the second contract, Lehman Brothers sells
5-year protection on Shell to British Airways on January 5, 2006.
In both contracts, British Airways computes the mark-to-market value
of the contract on May 1, 2008. We consider
different correlation scenarios among the three names.
The CDS quotes of the three names on those dates are reported
in Tables \ref{tab:spreads_2006} and \ref{tab:spreads_2008}.

\begin{table}[h!]
\begin{center}
\begin{tabular}{|c|c|c|c|}
\hline
Maturity & Royal Dutch Shell & Lehman Brothers  & British Airways
\\ \hline
\hline
1y & 4&6.8&10\\\hline
2y & 5.8&10.2&23.2\\\hline
3y & 7.8&14.4&50.6\\\hline
4y & 10.1&18.7&80.2\\\hline
5y & 11.7&23.2&110\\\hline
6y & 15.8&27.3.3&129.5\\\hline
7y & 19.4&30.5&142.8\\\hline
8y & 20.5&33.7&153.6\\\hline
9y & 21&36.5&162.1\\\hline
10y & 21.4&38.6&168.8\\\hline
\end{tabular}
\caption{Market spread quotes in basis points for Royal Dutch Shell,
  Lehman Brothers and British Airways on January 5, 2006.}
\label{tab:spreads_2006}
\end{center}
\end{table}

\begin{table}[h!]
\begin{center}
\begin{tabular}{|c|c|c|c|}
\hline
Maturity & Royal Dutch Shell & Lehman Brothers  & British Airways
\\ \hline
\hline
1y & 24&203&151\\\hline
2y & 24.6&188.5&230\\\hline
3y & 26.4&166.75&275\\\hline
4y & 28.5&152.25&305\\\hline
5y & 30&145&335\\\hline
6y & 32.1&136.3&342\\\hline
7y & 33.6&130&347\\\hline
8y & 35.1&125.8&350.6\\\hline
9y & 36.3&122.6&353.3\\\hline
10y & 37.2&120&355.5\\\hline
\end{tabular}
\caption{Market spread quotes in basis points for Royal Dutch Shell,
  Lehman Brothers and British Airways on May 1, 2008.}
\label{tab:spreads_2008}
\end{center}
\end{table}

The calibrated parameters for the CIR process dynamics on those dates are reported in
 Table \ref{tab:CIR_param_Lehm} and Table \ref{tab:CIR_param_Lehm2008} respectively.
The calibration is done assuming zero shift and inaccessibility of the origin.
To have a perfect calibration we should add a shift, but in this case the quality of fit is relatively good also without shift and this keeps the model simple. The absolute calibration errors $|S^{\mbox{\tiny CIR}}_{\mbox{\tiny CDS}} - S^{\mbox{\tiny MID-MKT}}_{\mbox{\tiny CDS}}|$ for the different names at the different dates range from less than one basis point to a maximum of 20 basis points for BA quotes about 355 basis points and of 23 for Lehman quotes about 188 basis points. These errors can be zeroed by introducing a shift, but for illustrating the CVA features the lack of shift does not compromise the analysis.

\begin{table}
\begin{center}
\begin{tabular}{|c|c|c|c||c|c|}
  \hline
  Credit Risk Levels (2006) & $y(0)$ & $\kappa$ & $\mu$ & $\nu$ \\
  \hline
  \hline
  Lehman Brothers (name ``0'')  & 0.0001 & 0.036 & 0.0432 & 0.0553 \\
  \hline
  Royal Dutch Shell (name ``1'') & 0.0001 & 0.0394 & 0.0219 &  0.0192 \\
  \hline
  British Airways (name ``2'')  & 0.00002 & 0.0266 & 0.2582 & 0.0003 \\
  \hline
\end{tabular}
\caption{The CIR parameters of Lehman Brothers, Royal Dutch Shell and
  British Airways calibrated to the market quotes of CDS on January 5,
  2006.}
\label{tab:CIR_param_Lehm}
\end{center}
\end{table}

\begin{table}
\begin{center}
\begin{tabular}{|c|c|c|c||c|c|}
  \hline
  Credit Risk Levels (2008) & $y(0)$ & $\kappa$ & $\mu$ & $\nu$ \\
  \hline
  \hline
  Lehman Brothers (name ``0'')  & 0.6611 & 7.8788 & 0.0208 &  0.5722 \\
  \hline
  Royal Dutch Shell (name ``1'') & 0.003 & 0.1835 & 0.0089 &  0.0057\\
  \hline
  British Airways (name ``2'')  & 0.00001 & 0.6773 & 0.0782 & 0.2242 \\
  \hline
\end{tabular}
\caption{The CIR parameters of Lehman Brothers, Royal Dutch Shell and
  British Airways calibrated to the market quotes of CDS on May 1,
  2008.}
\label{tab:CIR_param_Lehm2008}
\end{center}
\end{table}

The procedure used for mark-to-market value valuation is detailed next:

\begin{itemize}
\item [(a)] We take the CDS quotes of British Airways, Lehman Brothers and Royal Dutch Shell on January 5, 2006 and calibrate the parameters of the CIR  processes associated to the three names assuming zero shift and inaccessibility of the origin. The results obtained from the calibration are the ones reported in Table \ref{tab:CIR_param_Lehm}.

\item [(b)] We calculate the value of the five year
  risk-adjusted CDS contract starting at $T_{a}$ = January 5, 2006 and
  ending five years later at $T_{b}$ = January 5, 2011 as
\begin{equation}
CDS_{a,b}^D(T_a,S_1,\TLGD{0}{1}{2}) =
CDS_{a,b}(T_a,S_1,\LGD{1}) - \brcvacds_{a,b}(T_a,S_1,\TLGD{0}{1}{2})
\label{eq:CDS_adj}
\end{equation}
where $S_1 = 120 $ bps is the five-year spread quote of Royal Dutch Shell at time $T_a$, $\cds_{a,b}(T_a,S_1,\LGD{1})$ is the value of
the equivalent CDS contract which does not account for counterparty
 risk given by Eq. (\ref{eq:cds_payoff}) and the loss given
 default of the three names are taken from a market provider and equal to 0.6.
\item[(c)]
Let $T_{c}$ = May 1, 2008, be the time at which British Airways calculates the
mark to market value of the CDS contract. We keep the CIR parameters of
British Airways and Royal Dutch Shell at the
same values calibrated in (a). We vary the volatility of the CIR
process associated to Lehman Brothers, while keeping the other
parameters fixed. We take the market CDS quotes of Lehman Brothers, British
Airways and Royal Dutch Shell and recompute the shift process as
\begin{eqnarray}
\nonumber \Psi_i(t;\bm{\beta}_i) &=&   \log \left(
\frac{\mathbb{E}[e^{-Y_i(t)}\
]}{\mathbb{Q}(\tau_i >t)_{market}} \right) \\
&=& \log \left(\frac{P^{CIR}(0,t,\bm{\beta}_i)}{\mathbb{Q}(\tau_i
  >t)_{market}}\
\right)
\label{eq:shift_integrated_Lehm}
\end{eqnarray}
for any $T_{c} < t < T_{d}$, where $T_{d}$ = May 1, 2013. Here the
market survival probabilities $\mathbb{Q}(\tau_i  >t)_{market}$, are
stripped from the CDS quotes of the three names at the date, May, 1,
2008, using a piece-wise linear hazard rate function. We compute the value $\cds_{c,d}^D(T_c,S_1,\LGD{1})$ of a risk
adjusted CDS contract starting at $T_{c}$ and maturing at $T_{d}$,
where the five year running spread premium as well as the loss given
defaults of the three parties are the same as in $T_a$. We have
\begin{equation}
\cds^D_{c,d}(T_{c},S_1,\TLGD{0}{1}{2}) = \cds_{c,d}(T_c,S_1,\LGD{1}) -
\brcvacds_{c,d}(T_c,S_1,\TLGD{0}{1}{2})
\label{eq:adj}
\end{equation}
where $\cds_{c,d}(T_c,S_1,\LGD{1})$ is the value of the equivalent
CDS contract which does not account for counterparty risk and
$\brcvacds_{c,d}(T_c, S_1,\LGD{1})$ is the adjustment for the period
$[T_c,T_d]$ calculated at time $T_c$.

\item [(d)] We calculate the mark-to-market value of the CDS contract as follows:
\begin{equation}
MTM_{a,c}(S_1,\TLGD{0}{1}{2})  = CDS^D_{c,d}(T_c,S_1,\TLGD{0}{1}{2}) -
 \frac{CDS^D_{a,b}(T_a,S_1,\TLGD{0}{1}{2})}{D(T_a,T_c)}
\end{equation}
\end{itemize}

Table \ref{tab:CIR_param_Lehm_invest} reports the MTM value of the CDS
  contract between British Airways and Lehman Brothers on default of Royal Dutch Shell under a number of
  correlation scenarios. The CDS contract agreed
  on January 5, 2006 is marked to market on May, 1, 2008 by British
  Airways using the four-step procedure described above. We check the
  effect of the increasing riskiness of Lehman Brothers by varying the
  volatility of the CIR process associated to Lehman.
\begin{table}[h!]
\begin{center}
\begin{scriptsize}
\begin{tabular}{|l|l|c|c|c|c|c|c|}
\hline
\textbf{($r_{01},r_{02},r_{12}$)}&\textbf{Vol. parameter $\nu_1$}&\textbf{0.01}&\textbf{0.10}&\textbf{0.20}&\textbf{0.30}&\textbf{0.40}&\textbf{0.50}\\
& \textbf{CDS Impled vol}&\textbf{1.5\%}&\textbf{15\%}&\textbf{28\%}&\textbf{37\%}&\textbf{42\%}&\textbf{42\%}\\
 \hline
 \textbf{(-0.3, -0.3, 0.6)}&(LEH Pay, BAB Rec)&39.1(2.1)&44.7(2.0)&51.1(1.9)&58.4(1.4)&60.3(1.7)&63.8(1.1)\\
&(BAB Pay, LEH Rec)&-84.2(0.0)&-83.8(0.1)&-83.5(0.1)&-83.8(0.1)&-83.8(0.2)&-83.8(0.2)\\
\hline
\textbf{(-0.3, -0.3, 0.8)}&(LEH Pay, BAB Rec)&13.6(3.6)&22.6(3.2)&35.2(2.6)&43.7(2.0)&45.3(2.4)&52.0(1.4)\\
&(BAB Pay, LEH Rec)&-84.2(0.0)&-83.9(0.1)&-83.6(0.1)&-83.9(0.1)&-83.9(0.2)&-83.8(0.2)\\
\hline
\textbf{(0.6, -0.3, -0.2)}&(LEH Pay, BAB Rec)&83.1(0.0)&81.9(0.2)&81.6(0.3)&82.4(0.3)&82.6(0.3)&82.8(0.4)\\
&(BAB Pay, LEH Rec)&-55.6(1.8)&-58.7(1.7)&-66.1(1.4)&-71.3(1.1)&-73.2(1.0)&-74.1(0.9)\\
\hline
\textbf{(0.8, -0.3, -0.3)}&(LEH Pay, BAB Rec)&83.9(0.0)&82.9(0.1)&82.3(0.3)&82.9(0.2)&82.9(0.3)&83.0(0.3)\\
&(BAB Pay, LEH Rec)&-36.4(3.3)&-41.9(3.0)&-55.9(2.2)&-63.4(1.6)&-65.8(1.5)&-66.4(1.5)\\
\hline
\textbf{(0, 0, 0.5)}&(LEH Pay, BAB Rec)&50.6(1.5)&54.3(1.5)&59.2(1.5)&64.4(1.1)&65.5(1.3)&68.8(0.8)\\
&(BAB Pay, LEH Rec)&-80.9(0.2)&-80.5(0.3)&-80.9(0.4)&-82.3(0.3)&-82.6(0.3)&-82.8(0.3)\\
\hline
\textbf{(0, 0, 0.8)}&(LEH Pay, BAB Rec)&12.3(3.5)&21.0(3.0)&34.9(2.5)&41.3(2.1)&44.6(1.9)&50.6(1.4)\\
&(BAB Pay, LEH Rec)&-80.9(0.2)&-81.5(0.2)&-81.9(0.3)&-81.9(0.4)&-82.1(0.4)&-82.7(0.3)\\
\hline
\textbf{(0, 0, 0)}&(LEH Pay, BAB Rec)&78.1(0.2)&77.9(0.3)&79.5(0.5)&79.5(0.5)&80.1(0.6)&82.1(0.4)\\
&(BAB Pay, LEH Rec)&-81.6(0.2)&-81.9(0.2)&-82.3(0.3)&-82.2(0.4)&-82.7(0.3)&-83.2(0.3)\\
\hline
\textbf{(0, 0.7, 0)}&(LEH Pay, BAB Rec)&77.3(0.3)&77.3(0.4)&78.5(0.5)&79.2(0.5)&79.7(0.6)&81.5(0.4)\\
&(BAB Pay, LEH Rec)&-81.2(0.2)&-81.8(0.2)&-81.9(0.3)&-80.8(1.3)&-82.4(0.3)&-82.6(0.3)\\
\hline
\textbf{(0.3, 0.2, 0.6)}&(LEH Pay, BAB Rec)&54.1(1.4)&56.7(1.3)&62.5(1.1)&63.6(1.1)&66.4(0.9)&69.7(0.6)\\
&(BAB Pay, LEH Rec)&-81.3(0.2)&-81.7(0.2)&-81.4(0.4)&-81.3(0.5)&-81.6(0.4)&-82.1(0.4)\\
\hline
\textbf{(0.3, 0.3, 0.8)}&(LEH Pay, BAB Rec)&22.8(4.2)&28.8(3.5)&38.6(2.9)&42.6(2.9)&45.9(2.5)&52.0(2.2)\\
&(BAB Pay, LEH Rec)&-83.0(0.2)&-83.2(0.2)&-82.8(0.3)&-82.4(0.4)&-82.5(0.4)&-82.9(0.4)\\
\hline
\textbf{(0.5, 0.5, 0.5)}&(LEH Pay, BAB Rec)&62.8(0.8)&64.5(0.8)&67.7(0.8)&68.5(0.9)&71.3(0.7)&73.2(0.6)\\
&(BAB Pay, LEH Rec)&-67.4(1.1)&-70.4(0.9)&-72.9(0.9)&-74.4(0.9)&-75.8(0.8)&-76.7(0.7)\\
\hline
\textbf{(0.7, 0, 0)}&(LEH Pay, BAB Rec)&77.4(0.2)&77.3(0.3)&78.9(0.5)&79.1(0.5)&79.9(0.5)&81.4(0.4)\\
&(BAB Pay, LEH Rec)&-47.3(2.2)&-55.0(1.9)&-61.6(1.6)&-65.0(1.5)&-67.5(1.3)&-69.6(1.1)\\
\hline
\end{tabular}
\end{scriptsize}
\caption{Value of the CDS contract between British Airways and
  Lehman Brothers on default of Royal Dutch Shell agreed on January
  5, 2006 and marked to market by Lehman Brothers on
  May 1, 2008. The pairs (LEH Pay, BAB Rec) and (BAB Pay, LEH Rec)
denote respectively the mark-to-market value when Lehman Brothers is
  the CDS receiver and CDS payer. The mark-to-market value of the CDS
  contract without risk adjustment when Lehman Brothers is
  respectively payer (receiver) is 84.2(-84.2) bps, due to the
  widening of the CDS spread curve of Royal Dutch Shell.
}
\label{tab:CIR_param_Lehm_invest}
\end{center}
\end{table}
Table \ref{tab:CIR_param_Lehm_invest} reveals sensitivity of the
BR-CVA to both default correlation and credit risk volatility of Lehman. We notice the following behavior.
If British Airways is negatively correlated or uncorrelated with Royal Dutch Shell, see triples
$(0.6,-0.3,-0.2)$, $(0.8,-0.3,-0.3)$, and $(0.7,0,0)$, then the mark-to-market value of the
risk adjusted CDS contract appears to be the largest for Lehman (this is true both if Lehman is the CDS payer
and the CDS receiver).
%If Lehman is negatively correlated or uncorrelated with Royal Dutch Shell
%and British Airways, which are instead positively correlated with each other, then the mark-to-market value of the
%risk adjusted CDS contract appears to be the smallest for Lehman (this is true both if Lehman is the CDS payer
%and the CDS receiver). The risk-adjusted mark-to-market value, instead, appears to be the largest in scenarios
%where Lehman is positively correlated only with one of the entity, while these two are uncorrelated with each other.
The adjustments are quite sensitive to the credit spreads volatility of Royal Dutch Shell. In particular, it appears from Table \ref{tab:CIR_param_Lehm_invest} that increases in credit spreads volatility of Shell increase the mark-to-market valuation of the CDS contract when Lehman is the CDS payer and decrease the contract valuation when Lehman is the CDS receiver.
This is the case because larger credit spreads volatility increases the number of scenarios where the
counterparty British Airways precedes Shell in defaulting. When Lehman is the CDS payer,
this translates in larger CDS contract valuation for Lehman, as the negative adjustment done at the counterparty default time
if the option on the residual net present value is in the money (and this is the case as the default intensity of Shell in 2008 has increased) does not take place. Conversely, when Lehman is the CDS receiver, this implies smaller CDS contract valuations for Lehman due to a symmetric reasoning.
%This is because as the number of scenarios where Shell defaults earlier increases, then the out of the money option in the residual
%the receiver CDS contract for the receiver Lehman becomes less and less valuable as the CDS contract
%option on the residual net present value is out of the money and is exerci.

We next invert the role of Lehman Brothers and Royal Dutch Shell, hence
Royal Dutch Shell becomes the investor and Lehman Brothers the
reference entity.

We consider the following two CDS contracts. In the
first contract Royal Dutch Shell buys 5-year protection on Lehman from
British Airways on January 5, 2006. In the second contract, Royal
Dutch Shell sells 5-year protection on Lehman to British Airways on January 5, 2006.
As in the earlier case, British Airways computes the mark-to-market value
of the two CDS contracts on May 1, 2008.

%We consider again the effect of Lehman's volatility on the
%mark-to-market value of the CDS contract for British Airways on
%May 1, 2008 under the same correlation scenarios as above.
The results are reported in Table \ref{tab:CIR_param_Lehm_reference}.

\begin{table}[h!]
\begin{center}
\begin{scriptsize}
\begin{tabular}{|l|l|c|c|c|c|c|c|}
\hline
\textbf{($r_{01},r_{02},r_{12}$)}&\textbf{Vol. parameter $\nu_1$}&\textbf{0.01}&\textbf{0.10}&\textbf{0.20}&\textbf{0.30}&\textbf{0.40}&\textbf{0.40}\\
& \textbf{CDS Impled vol}&\textbf{1.5\%}&\textbf{15\%}&\textbf{28\%}&\textbf{37\%}&\textbf{42\%}&\textbf{42\%}\\
 \hline
\textbf{(-0.3, -0.3, 0.6)}&(BAB Pay, RDSPLC Rec)&513.0(1.9)&512.4(1.9)&512.8(1.9)&512.7(1.9)&513.4(1.9)&514.0(1.8)\\
&(RDSPLC Pay, BAB Rec)&-520.0(0.2)&-520.0(0.2)&-520.0(0.2)&-520.0(0.2)&-520.0(0.2)&-520.1(0.2)\\
\hline
\textbf{(-0.3, -0.3, 0.8)}&(BAB Pay, RDSPLC Rec)&511.1(2.3)&511.1(2.4)&511.4(2.3)&511.3(2.3)&511.9(2.3)&513.0(2.2)\\
&(RDSPLC Pay, BAB Rec)&-520.0(0.2)&-520.0(0.2)&-520.0(0.2)&-520.0(0.2)&-520.0(0.2)&-520.1(0.2)\\
\hline
\textbf{(0.6, -0.3, -0.2)}&(BAB Pay, RDSPLC Rec)&525.9(0.1)&525.9(0.1)&525.9(0.1)&525.9(0.1)&525.9(0.1)&525.8(0.1)\\
&(RDSPLC Pay, BAB Rec)&-442.2(3.6)&-442.7(3.5)&-442.6(3.5)&-442.6(3.5)&-442.3(3.5)&-441.2(3.6)\\
\hline
\textbf{(0.8, -0.3, -0.3)}&(BAB Pay, RDSPLC Rec)&526.5(0.0)&526.5(0.0)&526.5(0.0)&526.5(0.0)&526.5(0.0)&526.5(0.0)\\
&(RDSPLC Pay, BAB Rec)&-405.1(5.2)&-405.3(5.2)&-403.3(5.3)&-405.4(5.2)&-407.6(5.1)&-403.9(5.3)\\
\hline
\textbf{(0, 0, 0.5)}&(BAB Pay, RDSPLC Rec)&516.0(1.4)&515.7(1.4)&515.5(1.5)&516.3(1.3)&515.6(1.5)&516.7(1.3)\\
&(RDSPLC Pay, BAB Rec)&-503.7(0.8)&-503.7(0.8)&-503.7(0.8)&-503.6(0.8)&-503.8(0.8)&-503.9(0.8)\\
\hline
\textbf{(0, 0, 0.8)}&(BAB Pay, RDSPLC Rec)&511.6(2.4)&511.5(2.4)&512.1(2.2)&512.1(2.2)&505.7(7.1)&512.6(2.2)\\
&(RDSPLC Pay, BAB Rec)&-503.7(0.8)&-503.7(0.8)&-503.7(0.8)&-507.5(4.0)&-503.7(0.8)&-503.9(0.8)\\
\hline
\textbf{(0, 0, 0)}&(BAB Pay, RDSPLC Rec)&524.2(0.3)&524.2(0.3)&524.1(0.3)&524.2(0.3)&524.2(0.3)&524.2(0.3)\\
&(RDSPLC Pay, BAB Rec)&-504.2(0.8)&-504.1(0.8)&-504.2(0.8)&-504.0(0.8)&-504.2(0.8)&-504.3(0.8)\\
\hline
\textbf{(0, 0.7, 0)}&(BAB Pay, RDSPLC Rec)&524.8(0.3)&524.7(0.4)&524.8(0.3)&524.7(0.3)&524.7(0.3)&524.8(0.3)\\
&(RDSPLC Pay, BAB Rec)&-504.3(0.8)&-504.3(0.8)&-504.4(0.8)&-504.2(0.8)&-504.4(0.8)&-504.5(0.8)\\
\hline
\textbf{(0.3, 0.2, 0.6)}&(BAB Pay, RDSPLC Rec)&516.6(1.5)&517.0(1.5)&516.6(1.5)&517.3(1.4)&517.1(1.5)&517.2(1.4)\\
&(RDSPLC Pay, BAB Rec)&-484.3(1.7)&-484.4(1.7)&-484.3(1.7)&-484.4(1.6)&-484.5(1.6)&-484.3(1.7)\\
\hline
\textbf{(0.3, 0.3, 0.8)}&(BAB Pay, RDSPLC Rec)&507.4(5.5)&505.6(5.7)&508.9(4.5)&508.1(3.5)&502.6(7.7)&497.5(14.2)\\
&(RDSPLC Pay, BAB Rec)&-487.0(6.0)&-484.5(1.7)&-490.6(4.9)&-492.7(9.4)&-487.1(2.7)&-488.3(3.3)\\
\hline
\textbf{(0.5, 0.5, 0.5)}&(BAB Pay, RDSPLC Rec)&519.6(1.1)&519.6(1.1)&519.1(1.2)&519.7(1.1)&519.7(1.1)&519.4(1.1)\\
&(RDSPLC Pay, BAB Rec)&-460.2(2.8)&-460.2(2.8)&-459.6(2.8)&-460.1(2.8)&-459.4(2.8)&-458.1(2.8)\\
\hline
\textbf{(0.7, 0, 0)}&(BAB Pay, RDSPLC Rec)&523.8(0.3)&523.8(0.3)&523.7(0.3)&523.8(0.3)&523.8(0.3)&523.9(0.3)\\
&(RDSPLC Pay, BAB Rec)&-426.3(4.3)&-426.0(4.3)&-426.5(4.3)&-427.5(4.2)&-428.8(4.2)&-424.4(4.4)\\
\hline
\end{tabular}
\end{scriptsize}
\caption{Value of the CDS contract between British Airways and
  Royal Dutch Shell on default of Lehman Brothers agreed on January
  5, 2006 and marked to market by British Airways on
  May 1, 2008. The pairs (RDSPLC Pay, BAB Rec) and (BAB Pay, RDSPLC Rec)
denote respectively the mark-to-market value when British Airways is
  the CDS receiver and CDS payer. The mark-to-market value of the CDS
  contract without risk adjustment when British Airways is respectively payer (receiver) is 529(-529) bps, due to the widening of the CDS spread curve of Lehman Brothers.}
\label{tab:CIR_param_Lehm_reference}
\end{center}
\end{table}

The risk-adjusted mark-to-market value of the CDS contract is sensitive to correlation  and
it follows a pattern similar to the one discussed earlier for the other CDS contract.
However, differently from the previous case, there is less sensitivity to credit spreads volatility.
As the default intensity of Lehman on May 1 2008 is already high, we have that increases in its credit spreads volatility
does not vary significantly the number of scenarios where Lehman is the first to default. Therefore, the
risk-adjusted mark-to-market value of the CDS contract does not depend much on credit spreads volatility in this case.

%This is because, the marginal increases of Lehman's default intensity with respect to its credit spreads volatility
%is smaller than the marginal increase of Royal Dutch Shell default intensity with respect to its credit spreads volatility.
%Therefore, the marginal increase in the number of scenarios where Lehman's default precedes BA's default
%is smaller that marginal increase in the number of scenarios where Shell's default precedes British Airways.
%The conclusion is that the variation in the number of adjustments computed in this second case (Lehman being the reference entity)
%as the credit spreads volatility of Lehman increases, is smaller and thus the mark-to-market value of the risk-adjusted CDS contract
%does not vary much.

\section{Conclusions}\label{sec:conclusion}
We have provided a general framework for calculating the bilateral
counterparty credit valuation adjustment (BR-CVA) for payoffs
exchanged between two parties, an investor and her counterparty. We have then specialized our analysis to the case where also the underlying portfolio is sensitive to a third credit event, and in particular to the case where the underlying portfolio is a credit default swap on a third entity. We have then developed a Monte-Carlo numerical scheme to
evaluate the formula and thus compute the BR-CVA in the specific case
of credit default swap contracts. We have provided a case study in Section \ref{sec:case_study}, and
experimented with different levels of credit risk and credit risk
volatilities of the three names as well as with different scenarios of
default correlation. The results obtained confirm that the adjustment
is sensitive to both default correlation and credit spreads volatility, having richly structured patterns that cannot be captured by rough multipliers. This points out that attempting to adapting the capital adequacy methodology  (Basel II) to evaluating wrong way risk by means of rough multipliers is not feasible. Our analysis confirms that also in the bilateral-symmetric case, wrong way risk - namely the supplementary risk that one undergoes when the correlation assumes the worst possible value - has a structured pattern that cannot be captured by simple multipliers applied to the zero correlation case.

\section{Acknowledgements}
The authors are grateful to Su Chen at Stanford University for critically reading the paper and pointing out some issues in the original derivation of the formula in Eq.~(\ref{eq:surv_prob_computation}).

\appendix
\section{Proof of the general counterparty risk pricing formula}\label{sec:proof_prop}
We next prove the proposition.
\proof
We have that
\begin{eqnarray}
\nonumber \Pi(t,T) &=& \cash(t, T) \\
&=& \mathbf{1}_{A \cup B}\cash(t, T) +
\mathbf{1}_{C \cup D}\cash(t, T) +   \mathbf{1}_{E \cup F}\cash(t, T)
\label{eq:default_free_portfolio}
\end{eqnarray}
since the events in Eq. (\ref{eq:event_set}) form a complete set.
From the linearity of the expectation, we can rewrite the right hand
side of Eq. (\ref{generalprice}) as
\begin{equation}
 \mathbb{E}_t \{\Pi(t,T) + \mathrm{LGD_0} \cdot \mathbf{1}_{A
 \cup B}  \cdot D(t,\tau_0) \cdot \left[ -
 \mathrm{NPV}(\tau_0)\right]^+
  - \mathrm{LGD_2} \cdot  \mathbf{1}_{C \cup D}
 \cdot D(t,\tau_2) \cdot \left[ \mathrm{NPV}(\tau_2)\right]^+\}
\label{general_linearity}
\end{equation}
We can then rewrite the formula in
Eq. (\ref{general_linearity}) using Eq.
\ref{eq:default_free_portfolio} as
\begin{eqnarray}
\nonumber &=& E_t[\mathbf{1}_{A \cup B} \cash(t, T) + (1 - \recinv) \mathbf{1}_{A \cup B} D(t,\tau_0)
       [-NPV(\tau_0)]^+ \\
\nonumber & & + \mathbf{1}_{C \cup D} \cash(t, T) + (\reccou-1) \mathbf{1}_{C \cup D} D(t,\tau_2)
       [NPV(\tau_2)]^+ \\
 & & +      \mathbf{1}_{E \cup F} \cash(t, T)] \nonumber \\
\nonumber &=& E_t[\mathbf{1}_{A \cup B} \cash(t, T) + (1 - \recinv) \mathbf{1}_{A \cup B} D(t,\tau_0)
       [-NPV(\tau_0)]^+] \\
\nonumber & & + E_t[\mathbf{1}_{C \cup D} \cash(t, T) + (\reccou-1) \mathbf{1}_{C \cup D} D(t,\tau_2)
       [NPV(\tau_2)]^+] \\
 & & +     E_t[\mathbf{1}_{E \cup F} \cash(t, T)]
\label{eq:decomposition_expectation}
\end{eqnarray}
We next develop each of the three expectations in the equality of
Eq. (\ref{eq:decomposition_expectation}).

The expression inside the first expectation can be rewritten as
\begin{eqnarray}
\nonumber & & \mathbf{1}_{A \cup B} \cash(t, T) + (1 - \recinv) \mathbf{1}_{A \cup B} D(t,\tau_0)
       [-NPV(\tau_0)]^+  \\
\nonumber &=&  \mathbf{1}_{A \cup B} \cash(t, T) + \mathbf{1}_{A \cup B} D(t,\tau_0)
       [-NPV(\tau_0)]^+ -  \recinv \mathbf{1}_{A \cup B} D(t,\tau_0) [-NPV(\tau_0)]^+ \\
\label{eq:default_investor}
\end{eqnarray}
Conditional on the information at $\tau_0$, the expectation of the
expression in Eq. (\ref{eq:default_investor}) is equal to
\begin{eqnarray}
\nonumber & & \hspace{-5mm}\mathbb{E}_{\tau_0}\left[\mathbf{1}_{A \cup B} \cash(t, T) + \mathbf{1}_{A \cup B} D(t,\tau_0)
       \left(-NPV(\tau_0)\right)^+ - \recinv \mathbf{1}_{A \cup B} D(t,\tau_0) [-NPV(\tau_0)]^+\right]  \\
\nonumber & = & \mathbb{E}_{\tau_0}[\mathbf{1}_{A \cup B} [\cash(t,
    \tau_0) + D(t,\tau_0) \cash(\tau_0,T) + D(t,\tau_0)
      \left(-\mathbb{E}_{\tau_0}\left[\cash(\tau_0,T)\right]\right)^+\\
\nonumber& & -\recinv  D(t,\tau_0) [-NPV(\tau_0)]^+]] \\
\nonumber &=& \mathbf{1}_{A \cup B} [\cash(t, \tau_0) + D(t,\tau_0)
       \mathbb{E}_{\tau_0}\left[\cash(\tau_0,T)\right] + D(t,\tau_0)
       \left(-\mathbb{E}_{\tau_0}\left[\cash(\tau_0,T)\right]\right)^+
        \\
\nonumber& & -\recinv D(t,\tau_0) [-NPV(\tau_0)]^+]\\
\nonumber &=& \mathbf{1}_{A \cup B} [\cash(t, \tau_0) +
       D(t,\tau_0)\left(\mathbb{E}_{\tau_0}\left[\cash(\tau_0,T)\right]\right)^+
-\recinv  D(t,\tau_0) [-NPV(\tau_0)]^+] \\
\nonumber &=& \mathbf{1}_{A \cup B} [\cash(t, \tau_0)
       + D(t,\tau_0)\left(NPV(\tau_0)\right)^+  -
\recinv  D(t,\tau_0) [-NPV(\tau_0)]^+]
\label{eq:terms_default_investor}
\end{eqnarray}
where the first equality in Eq. (\ref{eq:terms_default_investor})
follows because
\begin{equation}
\mathbf{1}_{A \cup B} \cash(t, T) = \mathbf{1}_{A \cup B} [\cash(t,
\tau_0) + D(t,\tau_0) \cash(\tau_0,T)]
\end{equation}
being the default time $\tau_0$ always smaller than $T$ under the event
$A \cup B$. Conditioning the obtained result on the information
available at $t$, and using the fact that $E_t[E_{\tau_0}[.]] =
E_t[.]$ due to $t < \tau_0$, we obtain
that the first term in Eq. (\ref{eq:decomposition_expectation}) is
given by
\begin{equation}
\mathbb{E}_{t}\left[\mathbf{1}_{A \cup B} \left[\cash(t, \tau_0) +
       D(t,\tau_0) (NPV(\tau_0))^+ - \recinv D(t,\tau_0)(-NPV(\tau_0))^+\right]\right]
\end{equation}
which coincides with the expectation of the third term in Eq. (\ref{generalpayoff}).

We next repeat a similar argument for the second expectation in
Eq. (\ref{eq:decomposition_expectation}). We have

\begin{eqnarray}
\nonumber & & \mathbf{1}_{C \cup D} \cash(t, T) + (\reccou-1) \mathbf{1}_{C \cup D} D(t,\tau_2)
       [NPV(\tau_2)]^+  \\
\nonumber &=&  \mathbf{1}_{C \cup D} \cash(t, T) - \mathbf{1}_{C \cup D} D(t,\tau_2)
       [NPV(\tau_2)]^+ +  \reccou \mathbf{1}_{C \cup D} D(t,\tau_2) [NPV(\tau_2)]^+ \\
\label{eq:default_counterparty}
\end{eqnarray}

Conditional on the information available at time $\tau_2$, we have
\begin{eqnarray}
\nonumber & & \hspace{-5mm}\mathbb{E}_{\tau_2}\left[\mathbf{1}_{C \cup D} \cash(t, T) - \mathbf{1}_{C \cup D} D(t,\tau_2)
       \left(NPV(\tau_2)\right)^+   \reccou \mathbf{1}_{C \cup D} D(t,\tau_2) [NPV(\tau_2)]^+ \right]  \\
\nonumber & = & \mathbb{E}_{\tau_2}[\mathbf{1}_{C \cup D}[\cash(t,
    \tau_2) + D(t,\tau_2) \cash(\tau_2,T) - D(t,\tau_2)
      \left(\mathbb{E}_{\tau_2}\left[\cash(\tau_2,T)\right]\right)^+\\
\nonumber & & +  \reccou  D(t,\tau_2)
       [NPV(\tau_2)]^+ ]]\\
\nonumber &=& \mathbf{1}_{C \cup D} [\cash(t, \tau_2) + D(t,\tau_2)
       \mathbb{E}_{\tau_2}\left[\cash(\tau_2,T)\right] - D(t,\tau_2)
       \left(\mathbb{E}_{\tau_2}\left[\cash(\tau_2,T)\right]\right)^+\\
\nonumber & & +  \reccou  D(t,\tau_2) [NPV(\tau_2)]^+]
         \\
\nonumber &=& \mathbf{1}_{C \cup D} \left[\cash(t, \tau_2) -
       D(t,\tau_2)\left(\mathbb{E}_{\tau_2}\left[-\cash(\tau_2,T)\right]\right)^+
+  \reccou  D(t,\tau_2) [NPV(\tau_2)]^+\right]\\
 &=& \mathbf{1}_{C \cup D} \left[\cash(t, \tau_2) -
       D(t,\tau_2)(-NPV(\tau_2))^+
+  \reccou  D(t,\tau_2) [NPV(\tau_2)]^+
\right]
\label{eq:terms_default_counterparty}
\end{eqnarray}
where the first equality follows because
\begin{equation}
\mathbf{1}_{C \cup D} \cash(t, T) = \mathbf{1}_{C \cup D} [\cash(t,
\tau_2) + D(t,\tau_2) \cash(\tau_2,T)]
\end{equation}
being the default time $\tau_2$ always smaller than $T$ under the event
$C \cup D$. Conditioning the obtained result on the information
available at $t < \tau_2$, we obtain
that the second term in Eq. (\ref{eq:decomposition_expectation}) is
given by
\begin{equation}
\mathbb{E}_{t}\left[\mathbf{1}_{C \cup D} \left[\cash(t, \tau_2) +
       D(t,\tau_2)  \reccou (NPV(\tau_2))^+ - D(t,\tau_2) (-NPV(\tau_2))^+
       \right]\right]
\end{equation}
which coincides exactly with the expectation of the second term in Eq. (\ref{generalpayoff}).

The third expectation in Eq. (\ref{eq:decomposition_expectation}) coincides with the
first term in Eq. (\ref{generalpayoff}), therefore
their expectations ought to be the same. Since we have proven that the
expectation of each term in Eq. (\ref{generalpayoff}) equals the
expectation of the corresponding term in
Eq. (\ref{eq:decomposition_expectation}), the desired result is
obtained.
\endproof

\section{Proof of the survival probability formula}\label{sec:surv_prob_proof}
\proof We have
\begin{eqnarray}
\nonumber & & \mathbf{1}_{C \cup D} \mathbf{1}_{\tau_1 > \tau_2} \mathbb{Q}(\tau_1 > t |
\mathcal{G}_{\tau_2})  \\
%\nonumber & = &
\nonumber & & = \mathbf{1}_{\tau_2 \leq T} \mathbf{1}_{\tau_2 \leq \tau_0} \left(\mathbf{1}_{t < \tau_2 < \tau_1} +
\mathbf{1}_{\tau_2 < t} \mathbf{1}_{\tau_1 \geq \tau_2}
\mathbb{E}[\mathbb{Q}(\Lambda_1(t) < \xi_1|\mathcal{G}_{\tau_2},\xi_1) | \mathcal{G}_{\tau_2} ] \right)  \\
\nonumber & & =\mathbf{1}_{\tau_2 \leq T}  \mathbf{1}_{\tau_2 \leq \tau_0} \mathbf{1}_{\bar A} +
\mathbb{E}[\mathbf{1}_{\tau_2 < t} \mathbf{1}_{\tau_1 \geq \tau_2} \mathbf{1}_{\tau_0 \geq \tau_2} \mathbb{Q}(\Lambda_1(t) < \xi_1|\mathcal{G}_{\tau_2},\xi_1) | \mathcal{G}_{\tau_2}]  \\
\nonumber & & = \mathbf{1}_{\tau_2 \leq T} \mathbf{1}_{\tau_2 \leq \tau_0} \left(\mathbf{1}_{\bar A} + \mathbf{1}_{\tau_2 < t} \mathbf{1}_{\tau_1 \geq \tau_2}
\mathbb{E}[F_{\Lambda_1(t) - \Lambda_1(\tau_2)}(\xi_1 -
\Lambda_1(\tau_2)) | \mathcal{G}_{\tau_2}, \{\tau_1 > \tau_2\}, \{\tau_0 > \tau_2\}] \right)  \\
\label{eq:first_step_proof}
\end{eqnarray}
The last step follows from the fact that
the $\Lambda_1(t) < \xi_1$ is the same as $\Lambda_1(t) -
\Lambda_1(\tau_2)  < \xi_1 - \Lambda_1(\tau_2)$ and the right hand
side $\xi_1 - \Lambda_1(\tau_2)$ becomes known once we condition on $\xi_1$ and
$\mathcal{G}_{\tau_2}$. Here, by $F_{\Lambda_1(t) -
  \Lambda_1(\tau_2)}$, we indicate the cumulative distribution
function of the integrated (shifted) CIR process $\Lambda_1(t) -
  \Lambda_1(\tau_2)$. Let us denote $\overline{U}_{i,j} = 1-\exp(-\Lambda_i(\tau_j))$,
where $0 \leq i,j \leq 3$ denote the three names under
consideration. Since $\xi_1 = -\log(1-U_1)$ and $\tau_1 = \Lambda_1^{-1}(\xi_1)$, we can rewrite
  the inner term in Eq. (\ref{eq:first_step_proof}) as
\begin{eqnarray}
\nonumber  & & \hspace{-1cm} \mathbf{1}_{\bar A} + \mathbf{1}_{\tau_2 < t} \mathbf{1}_{\tau_1 \geq
  \tau_2} \mathbb{E}[F_{\Lambda_1(t) - \Lambda_1(\tau_2)}(-\log(1-U_1)
  - \Lambda_1(\tau_2))|\mathcal{G}_{\tau_2}, \{\xi_1 >
  \Lambda_1(\tau_2)\}, \{\xi_0 > \Lambda_0(\tau_2)\}] \\
%\nonumber &=& \mathbf{1}_{\bar A} + \mathbf{1}_{\tau_2 < t} \mathbf{1}_{\tau_1 \geq
%  \tau_2} \mathbb{E}[F_{\Lambda_1(t) - \Lambda_1(\tau_2)}(-\log(1-U_1)
%  - \Lambda_1(\tau_2))|\mathcal{G}_{\tau_2}, \{-\log(1-U_1) >
%  \Lambda_1(\tau_2)\}, \{-\log(1-U_0) > \Lambda_0(\tau_2)\}]  \\
\nonumber &=&\mathbf{1}_{\bar A} + \mathbf{1}_{\tau_2 < t} \mathbf{1}_{\tau_1 \geq
  \tau_2} \mathbb{E}[F_{\Lambda_1(t) - \Lambda_1(\tau_2)}(-\log(1-U_1)
  - \Lambda_1(\tau_2))|\mathcal{G}_{\tau_2}, U_1 > \overline{U}_{1,2}, U_0 > \overline{U}_{0,2}]\\
\label{eq:second_step_proof}
\end{eqnarray}

 Then we can rewrite Eq. (\ref{eq:second_step_proof}) as
\begin{eqnarray}
\nonumber  & & \hspace{-1.5cm}\mathbf{1}_{\bar A} + \mathbf{1}_{\tau_2 < t} \mathbf{1}_{\tau_1 \geq
  \tau_2} \int_{0}^1 F_{\Lambda_1(t) - \Lambda_1(\tau_2)}(-\log(1-u_1)
  - \Lambda_1(\tau_2)) d\mathbb{Q}(U_1 < u_1 | \mathcal{G}_{\tau_2}, U_1 >
  \overline{U}_{1,2}, U_0 > \overline{U}_{0,2}) \\
%\nonumber &=&\mathbf{1}_{\bar A} + \mathbf{1}_{\tau_2 < t}
%\mathbf{1}_{\tau_1 \geq
%  \tau_2}  \int_{0}^1 F_{\Lambda_1(t)}(-\log(1-U_1)) d\mathbb{Q}(U_1 <
%  u_1 | \mathcal{G}_{\tau_2}, U_1 > \overline{U}_{1,2}, U_0 > \overline{U}_{0,2}) \\
\nonumber &=&\mathbf{1}_{\bar A} + \mathbf{1}_{\tau_2 < t} \mathbf{1}_{\tau_1
\geq
  \tau_2} \int_{\overline{U}_{1,2}}^1 F_{\Lambda_1(t)- \Lambda_1(\tau_2)}(-\log(1-u_1) - \Lambda_1(\tau_2)) d\mathbb{Q}(U_1 <
  u_1 | \mathcal{G}_{\tau_2}, U_1 > \overline{U}_{1,2}, U_0 > \overline{U}_{0,2}) \\
\end{eqnarray}
The conditional distribution may be computed as follows. Denote

\begin{equation}
C_{1|0,2}(u_1;U_2) := \mathbb{Q}(U_1 < u_1 | \mathcal{G}_{\tau_2},
U_1 > \overline{U}_{1,2}, U_0 > \overline{U}_{0,2})
\label{eq:cond_copula_dev}
\end{equation}
%As $U_1$ depends on $\mathcal{G}_{\tau_2}$ only through $U_2$, then
Eq. (\ref{eq:cond_copula_dev}) may be rewritten as

\begin{eqnarray}
\nonumber C_{1|0,2}(u_1;U_2) &=& \mathbb{Q}(U_1 < u_1 | U_2, U_1 > \overline{U}_{1,2}, U_0 > \overline{U}_{0,2}) \\
&=& \frac{\mathbb{Q}(U_1 < u_1, U_1 > \overline{U}_{1,2} | U_2, U_0
> \overline{U}_{0,2})}{\mathbb{Q}(U_1 > \overline{U}_{1,2} |U_2,U_0
> \overline{U}_{0,2})}
\label{eq:cond_copula}
\end{eqnarray}
Notice that $U_2$ is explicitly known once we condition on
$\mathcal{G}_{\tau_2}$ and is given by $1 -
\exp(\Lambda_2(\tau_2))$. Let us next express
Eq.~(\ref{eq:cond_copula}) in terms of the copula function. We start
with the numerator, which may be expressed as
\begin{eqnarray}
\nonumber \mathbb{Q}(U_1 < u_1, U_1 > \overline{U}_{1,2} | U_2, U_0
> \overline{U}_{0,2}) &=& \frac{\mathbb{Q}(U_1 < u_1, U_1 >
\overline{U}_{1,2}, U_0 > \overline{U}_{0,2}|U_2)}{\mathbb{Q}(U_0 > \overline{U}_{0,2} | U_2)} \\
\nonumber &=& \frac{\mathbb{Q}(U_1 < u_1, U_0 > \overline{U}_{0,2} |
U_2) - \mathbb{Q}(U_1 < \overline{U}_{1,2}, U_0 > \overline{U}_{0,2}
|
U_2)}{\mathbb{Q}(U_0 > \overline{U}_{0,2} | U_2)} \\
\label{eq:numdecomp}
\end{eqnarray}
We have that
\begin{eqnarray}
\nonumber \mathbb{Q}(U_1 < u_1, U_0 > \overline{U}_{0,2} | U_2) &=&
\mathbb{Q}(U_1 < u_1 | U_2) - \mathbb{Q}(U_1 < u_1, U_0 < \overline{U}_{0,2} | U_2) \\
&=& \frac{\partial C_{1,2}(u_1,u_2)}{\partial u_2}\bigg|_{u_2 = U_2}
- \frac{\partial C(\overline{U}_{0,2},u_1,u_2)}{\partial
u_2}\bigg|_{u_2 = U_2}
\end{eqnarray}
where $C_{1,2}$ denotes the bivariate copula connecting the default
times of names ``1'' and ``2'', while $C$ denotes the trivariate
copula. Similarly, we have that
\begin{eqnarray}
\mathbb{Q}(U_1 < \overline{U}_{1,2}, U_0 > \overline{U}_{0,2} | U_2)
&=& \frac{\partial C_{1,2}(\overline{U}_{1,2},u_2)}{\partial
u_2}\bigg|_{u_2 = U_2} - \frac{\partial
C(\overline{U}_{0,2},\overline{U}_{1,2},u_2)}{\partial
u_2}\bigg|_{u_2 = U_2}
\end{eqnarray}
\begin{comment}
The denominator in Eq.~(\ref{eq:numdecomp}) may be written as
\begin{eqnarray}
\mathbb{Q}(U_0 > \overline{U}_{0,2} | U_2) &=& 1 - \frac{\partial
C_{0,2} (\overline{U}_{0,2},u_2)}{\partial u_2}\bigg|_{u_2 = U_2}
\end{eqnarray}
where $C_{0,2}$ is the copula connecting the default times of name
``0'' and name ``2''.
\end{comment}
The denominator in Eq.~(\ref{eq:cond_copula})
may be computed as
\begin{eqnarray}
\mathbb{Q}(U_1 > \overline{U}_{1,2} | U_2, U_0 > \overline{U}_{0,2})
&=& \frac{\mathbb{Q}(U_1 > \overline{U}_{1,2}, U_0 >
\overline{U}_{0,2} | U_2)}{\mathbb{Q}(U_0 > \overline{U}_{0,2} |
U_2)} \label{eq:denominator}
\end{eqnarray}
where
\begin{eqnarray}
\nonumber \mathbb{Q}(U_1 > \overline{U}_{1,2}, U_0 >
\overline{U}_{0,2}| U_2) &=& \mathbb{Q}(U_0 >
\overline{U}_{0,2}|U_2) - \mathbb{Q}(U_0 >
\overline{U}_{0,2}, U_1 < \overline{U}_{1,2} |U_2) \\
\nonumber &=& 1-\frac{\partial C_{0,2}
(\overline{U}_{0,2},u_2)}{\partial u_2}\bigg|_{u_2 = U_2} -
\frac{\partial C_{1,2} (\overline{U}_{1,2},u_2)}{\partial
u_2}\bigg|_{u_2 = U_2} + \frac{\partial
C(\overline{U}_{0,2},\overline{U}_{1,2},u_2)}{\partial
u_2}\bigg|_{u_2 =
U_2}\\
\end{eqnarray}
\begin{comment}
while
\begin{eqnarray}
\mathbb{Q}(U_0 > \overline{U}_{0,2} | U_2) = 1 - \frac{\partial
C_{0,2} (\overline{U}_{1,2},u_2)}{\partial u_2}\bigg|_{u_2 = U_2}
\end{eqnarray}
\end{comment}
All together, we have that
\begin{eqnarray}
\nonumber C_{1|0,2}(u_1;U_2) &=& \frac{\frac{\partial C_{1,2}
(u_1,u_2)}{\partial u_2}\bigg|_{u_2 = U_2} - \frac{\partial C
(\overline{U}_{0,2},u_1,u_2)}{\partial u_2}\bigg|_{u_2 = U_2} -
\frac{\partial C_{1,2} (\overline{U}_{1,2},u_2)}{\partial
u_2}\bigg|_{u_2 = U_2} + \frac{\partial C
(\overline{U}_{0,2},\overline{U}_{1,2},u_2)}{\partial
u_2}\bigg|_{u_2 = U_2} }{1-\frac{\partial C_{0,2}
(\overline{U}_{0,2},u_2)}{\partial u_2}\bigg|_{u_2 = U_2} -
\frac{\partial C_{1,2}
(\overline{U}_{1,2},u_2)}{\partial u_2}\bigg|_{u_2 = U_2} + \frac{\partial
C(\overline{U}_{0,2},\overline{U}_{1,2},u_2)}{\partial
u_2}\bigg|_{u_2 =
U_2}} \\
\end{eqnarray}

\begin{comment}
The expression in Eq. (\ref{eq:cond_copula}) may be computed in
closed form for the Gaussian copula and are given \cite{Schon03} by

\begin{equation}
\frac{\partial C(u,v)}{\partial u} = \Phi \left(\frac{x- r_{1,2}
y}{\sqrt{1 - r_{1,2}^2}}\right) \label{eq:gauss_copula}
\end{equation}
where $x = \Phi^{-1}(u)$ and $y=\Phi^{-1}(v)$, and $r_{1,2}$ is the
correlation between the reference credit and the counterparty.
\end{comment}

\endproof
The case when the investor is the first to default is symmetric, leading to the conditional distribution
given by
\begin{eqnarray}
\nonumber C_{1|2,0}(u_1;U_0) &:=& \mathbb{Q}(U_1 < u_1 | U_2,
\{\tau_1 > \tau_0\},\{\tau_2 > \tau_0\}) \\
\nonumber &=& \mathbb{Q}(U_1 < u_1 | U_0,
U_1 > \overline{U}_{1,0}, U_0 > \overline{U}_{2,0}) \\
\nonumber &=& \frac{\frac{\partial C_{0,1} (u_0,u_1)}{\partial
u_0}\bigg|_{u_0 = U_0} - \frac{\partial C
(u_0,u_1,\overline{U}_{2,0})}{\partial u_0}\bigg|_{u_0 = U_0} -
\frac{\partial C_{0,1} (u_0,\overline{U}_{1,0})}{\partial
u_0}\bigg|_{u_0 = U_0} + \frac{\partial C
(u_0,\overline{U}_{1,0},\overline{U}_{2,0})}{\partial
u_0}\bigg|_{u_0 = U_0}}{1-\frac{\partial C_{0,2}
(u_0,\overline{U}_{2,0})}{\partial u_0}\bigg|_{u_0 = U_0} -
\frac{\partial C_{0,1}
(u_0,\overline{U}_{1,0})}{\partial u_0}\bigg|_{u_0 = U_0} + \frac{\partial C
(u_0,\overline{U}_{1,0},\overline{U}_{2,0})}{\partial
u_0}\bigg|_{u_0 = U_0}} \\
\end{eqnarray}
and the survival probability conditioned on the information known at the investor default time is given by
\begin{eqnarray}
\nonumber & & \mathbf{1}_{A \cup B} \mathbf{1}_{\tau_1 > \tau_0} \mathbb{Q}(\tau_1 > t |
\mathcal{G}_{\tau_0})  \\
\nonumber & & = \mathbf{1}_{\tau_0 \leq T} \mathbf{1}_{\tau_0 \leq \tau_2} \left(\mathbf{1}_{\bar B} + \mathbf{1}_{\tau_0 < t} \mathbf{1}_{\tau_1
\geq \tau_0} \int_{\overline{U}_{1,0}}^1 F_{\Lambda_1(t)- \Lambda_1(\tau_0)}(-\log(1-u_1) - \Lambda_1(\tau_0) ) dC_{1|2,0}(u_1;U_0)\right)  \\
\label{eq:surv_prob_computationapp}
\end{eqnarray}
\text{where}
\begin{equation}
\bar{B} = \{t < \tau_0 < \tau_1\}\\
\end{equation}

\end{document}